\documentclass[twocolumn,showpacs,reprint, preprintnumbers]{revtex4-2}
\usepackage{amsfonts}
\usepackage{amssymb}
\usepackage{amsmath}
\usepackage{graphicx}
\usepackage{dcolumn}
\usepackage{bm}
\usepackage{color}
\usepackage{braket}
\usepackage[colorlinks=true, allcolors=blue]{hyperref}
\usepackage{float}
\usepackage{appendix}
\usepackage{xcolor}
\usepackage{etoolbox}

\begin{document}
	
	\title{Emergence of Rich Dissipative Phases in the Anisotropic Quantum Rabi Model Driven by the $A^2$ Term}
	\author{Jun-Ling Wang$^{1}$ }
       \author{Yi-Bo Liu$^{1}$ }
	\author{Qing-Hu Chen$^{1,2,3,}$}
	\email{qhchen@zju.edu.cn}
	\affiliation{$^{1}$ Zhejiang Key Laboratory of Micro-Nano Quantum Chips and Quantum
		Control, School of Physics, Zhejiang University, Hangzhou 310027, China\\
        $^{2}$ Institute for Advanced Study in Physics, Zhejiang University, Hangzhou 310027, China \\
		$^{3}$ Collaborative Innovation Center of Advanced Microstructures, Nanjing
		University, Nanjing 210093, China}
	\date{\today}
	
	\begin{abstract}
	The open anisotropic quantum Rabi model (QRM) is studied in this work, with an explicit $A^2$ term incorporated. It is shown that, in the presence of cavity loss, anisotropy allows the steady-state superradiant transition to occur even when the $A^2$ term prohibits the transition in the isotropic case. The $A^2$ term reshapes and enriches the known steady-state phase diagram of the open anisotropic QRM, leading to an asymmetric structure composed of normal, superradiant, and bistable phases. In particular, the nonequilibrium tricritical points are generally separated from the merger point of the two critical branches that bound the superradiant region. Isolated bistable regions also emerge, and the number and positions of tricritical points are reduced. Notably, the anomalous photon-number fluctuation scaling is associated with the merger of the two critical branches, rather than with the tricritical points. These results clarify the role of the $A^2$ term contribution in the steady-state  structure of the open anisotropic QRM and provide a refined extension of the previously known dissipative phase diagram, offering a useful theoretical reference for future studies of related open light--matter systems.

	\end{abstract}
	
	\maketitle
	
	\section{\label{Introduction}Introduction}
	
	The Dicke model  serves as a paradigmatic framework for collective light--matter interactions, exhibiting a superradiant phase transition into a macroscopically excited photonic state in the thermodynamic limit \cite{DM, DM_PT,DM_PT2,emary2003prl}. Experimental progress in circuit QED \cite{tniemczyk2010np,forn2010prl,yoshihara2016np}, trapped ions \cite{trap_ions,lmduan2021nc}, and cold atomic systems \cite{cold_atom} has enabled light--matter coupling to reach the ultra-strong and deep-strong regimes, stimulating extensive studies on the normal phase (NP) to superradiant phase (SRP) transition in few-body systems, including the QRM \cite{ashhab2013superradiance,hwang2015prl,QRM}, its anisotropic generalizations \cite{mxliu2017prl,shen2017pra,xychen}, and related models \cite{cuiti2014prl,keeling2019aqt,hwang2016prl,zhang2021prl}.
	
	Nevertheless, the realization of an equilibrium  superradiant phase transition  in the standard QRM is precluded by a fundamental no-go theorem, which arises from the Thomas--Reich--Kuhn (TRK) sum rule and the associated $A^2$ term \cite{A2original,keeling2007jpb}. This limitation has sparked sustained debate regarding gauge invariance, the validity of the two-level approximation, and the microscopic implementation of the $A^2$ term in circuit-QED and other platforms \cite{vukics2014prl,ciuti2010nc,viehmann2011prl,andolina2019prb,nazir2020prl,rouse2023theory}. To circumvent this constraint, a variety of extrinsic approaches have been proposed, such as employing three-level systems \cite{ciuti2013pra}, introducing direct interatomic couplings \cite{taoliu2012pla,ymwang2020pra}, utilizing antisqueezing protocols \cite{Peng}, and exploring multimode configurations \cite{ymwang2020pra}. More recently, anisotropy has been established as an effective mechanism for overcoming the no-go theorem constraints on the  superradiant phase transition~\cite{Dicke_Stark,ye2025PRA}.
	
	Since realistic quantum systems are inherently coupled to their environment, the study of open quantum dynamics is indispensable. Nonequilibrium dissipative phenomena have been widely explored in driven-dissipative photonic systems \cite{bartolo2016pra,krimer2019prl,le2013steady,le2014bose}, dissipative Rydberg systems \cite{carr2013nonequilibrium,marcuzzi2014universal}, and reservoir-engineered many-body systems \cite{diehl2008quantum,kraus2008preparation,verstraete2009quantum}. Also, Liouvillian criticality provides a spectral framework for characterizing dissipative phase transitions (DPTs) in nonequilibrium steady states through steady-state nonanalyticity and Liouvillian-gap closing \cite{minganti2018spectral, rossini2021pr}. Experimental examples include semiconductor microcavities \cite{rodriguez2017prl}, superconducting circuit-QED lattices \cite{fitzpatrick2017prx}, and optical-lattice Bose gases with dissipation-controlled critical behavior \cite{tomita2017sciadv}.
	
	In light--matter systems, steady-state DPT from NP to SRP has attracted considerable attention. The DPT of the Dicke model has been both theoretically studied \cite{torre2013keldysh,keeling2010collective} and experimentally realized in atomic Bose–Einstein condensates coupled to optical cavities \cite{brennecke2013pnas,klinder2015pnas,kollar2017nc}. In addition, the open QRM exhibits a dissipative phase transition in the frequency limit \cite{Hwang2018PRA}, providing a versatile platform for exploring dissipative quantum criticality in small, well-controlled quantum systems \cite{cai2022cpl,zhao2025prl}. Moreover, under tunable anisotropy, the open QRM displays a rich phase diagram, including first-order and multicritical DPTs \cite{lyu2024PERL}. In the ultrastrong-coupling regime, dressed-state master equations have been developed to consistently account for environment-induced transitions between the eigenstates of the light-matter interaction systems \cite{beaudoin2011dissipation,ridolfo2012photon}.
	
	Given that anisotropy in the closed QRM can modify the restriction imposed by the no-go theorem and reshape the ground-state phase diagram in the presence of the TRK-related $A^2$ term, it is natural to ask whether a similar mechanism can operate in the open system. This motivates our investigation of the dissipative anisotropic QRM with an explicit $A^2$ term. We first clarify whether the superradiant transition remains prohibited in the open isotropic case once the TRK-related $A^2$ term is included. We then examine whether anisotropy can sustain a steady-state DPT and how the $A^2$ term modifies the resulting phase boundaries, bistable regions, tricritical structure, and fluctuation scaling.
	
	This paper is structured as follows. Section \ref{sec2} introduces the anisotropic QRM with the $A^2$ term and underlying symmetry. In Sec. \ref{sec3}, we present the steady-state phase diagram derived from mean-field equations and perform the stability analysis, revealing the emergence of normal, superradiant, and bistable phases. Section \ref{sec4} develops effective master equations for both the normal and superradiant phases in the limit of infinite qubit–cavity frequency ratio, enabling the calculation of quantum fluctuations and the identification of universal critical exponents.  In Sec. \ref{sec6}, we complement the analysis with numerical simulations of spin dynamics and the Wigner function of the cavity field, providing a direct visualization of the steady states across different phases. Finally, we summarize our findings and discuss their implications in Sec. \ref{conclusion}. 

\section{ \label{sec2}Model and symmetry}

The anisotropic QRM with the $A^2$ term can be described as
\begin{equation}
\begin{aligned} H &= \omega a^\dagger a + \frac{\Delta}{2}\sigma_z +
D\left(a + a^\dagger\right)^2 \\
&+g\left[\left(a\sigma_{+} + a^\dagger
\sigma_{-}\right) + \tau \left(a^\dagger\sigma_{+} +
a\sigma_{-}\right)\right],
\end{aligned}  
\label{Hamiltonian1}
\end{equation}
where $a^{\dagger}(a)$ is the creation (annihilation) operator of the light field with frequency $\omega$. $\sigma_{\pm}=(\sigma_x\pm i\sigma_y)/2$ and $\sigma_z$ are Pauli operators of the two-level atom with transition frequency $\Delta$. Parameter $\tau$ characterizes the anisotropy between the
counter-rotating-wave (CRW) and rotating-wave (RW) terms, i.e., $g_{\rm cr}=\tau g_{\rm r}$ with $g_{\rm r}=g$. Thus, $\tau=1$
corresponds to the isotropic QRM, $\tau=0$ to the Jaynes--Cummings limit, and $\tau\neq1$ to an effective anisotropic coupling.
The parameter $\tau$ may be positive or negative, depending on the relative sign and phase of the engineered RW and CRW couplings. Such anisotropic interactions can be controlled through periodic frequency modulation in circuit QED \cite{wang2019simulating,qtxie2014prx},
magnetic-flux driving of superconducting flux qubits \cite{wang2018quantum}, trapped ions \cite{trap_ions}, and magnon--spin-qubit hybrid systems \cite{skogvoll2021pra,erlingsson2010prb}. Also, the anisotropic QRM preserves
the same $\mathbb{Z}_2$ symmetry as the standard QRM, described
by $P=-\sigma_z e^{i\pi a^\dagger a}$, even in the presence of the $\mathrm{A}^2$
term.

Here, $D$ denotes the strength of the $A^2$ term and is parameterized as $D=\kappa D_0$ with $D_0=g^2/\Delta$ and $\kappa\geq1$. Microscopically, the $A^2$ term originates from the quadratic contribution in the minimal-coupling Hamiltonian, $(\mathbf{p}-e\mathbf{A})^2/2m$ \cite{A2original,andolina2019prb}. The lower bound $\kappa=1$ represents the minimal diamagnetic contribution required by the TRK sum rule, while $\kappa>1$ may account for additional oscillator strengths, higher-energy transitions, or platform-dependent quadratic field terms. 

The quadratic field contribution associated with the $A^2$ term can be essential and cannot be neglected in ultrastrong light--matter coupling, including cavity and circuit QED \cite{A2original,garcia2015light,bamba2016superradiant},
trapped ions \cite{puebla2017protected}, and intersubband polariton systems in semiconductor quantum wells
\cite{ciuti2005quantum,todorov2012intersubband}.
Hamiltonian~\eqref{Hamiltonian1} is therefore treated as an effective framework for examining how the $A^2$ term and anisotropic light--matter coupling jointly shape the dissipative phase structure.

In an open quantum system, the dynamics of the open anisotropic QRM is governed by a Lindblad master equation \cite{Hwang2018PRA}
\begin{equation}
\dot{\rho} = -i\left[H,\rho\right] + \gamma \mathcal{D}\left[a\right],
\label{lindblad}
\end{equation}
where $\mathcal{D}[a] = 2 a \rho a^\dagger - a^\dagger a \rho - \rho
a^\dagger a$ is the dissipative superoperator with damping rate $\gamma$. The cavity loss is modeled by the Markovian dissipator $\gamma\mathcal D[a]$. This local form is not a general microscopic master equation for ultrastrong-coupling systems, where the cavity and qubit environments cannot in general be treated independently, but it can be effectively implemented in engineered platforms such as driven trapped ions \cite{Hwang2018PRA} or cavity-QED systems based on cavity-assisted Raman processes \cite{grimsmo2013cavity}. Thus, Eq.~\eqref{lindblad} provides an effective dissipative framework for studying the DPT in light-matter interaction systems \cite{lyu2024PERL,Hwang2018PRA,kirton2017suppressing,gelhausen2018dissipative}. For passive ultrastrong-coupling realizations, dressed-state or gauge-consistent dissipators may be required.

The cavity damping in Eq.~\eqref{lindblad} does not break the $\mathbb{Z}_{2}$ symmetry. Since there is no explicit driving field that would violate the symmetry, the open anisotropic QRM remains $\mathbb{Z}_{2}$-symmetric. For the open QRM without the $A^2$ term, the critical point is shifted by dissipation to $g_c=\sqrt{\omega \Delta}\sqrt{1+\gamma^{2}/\omega^{2}}/2$ as derived in Ref.~\cite{Hwang2018PRA}. In contrast, when $\tau = 0$, the QRM reduces to the JCM and no DPT occurs. The absence of critical behavior in this limit stems from the fact that the Hamiltonian contains only particle-number-conserving interactions, while dissipation continuously removes excitations from the system, eventually driving it into a trivial vacuum steady-state. 

\section{\label{sec3}steady-state phase diagram}

In this section, we first derive the steady-state phase diagram and then perform a stability analysis based on the Lindblad master equation Eq.~\eqref{lindblad} for the system described by Eq.~\eqref{Hamiltonian1}, under the limit $\Delta
/\omega \rightarrow \infty $ at the mean-field level.

\subsection{Mean-field dissipative phase transitions}

\begin{figure*}[tbp]
\centering
\includegraphics[width=0.95\textwidth]{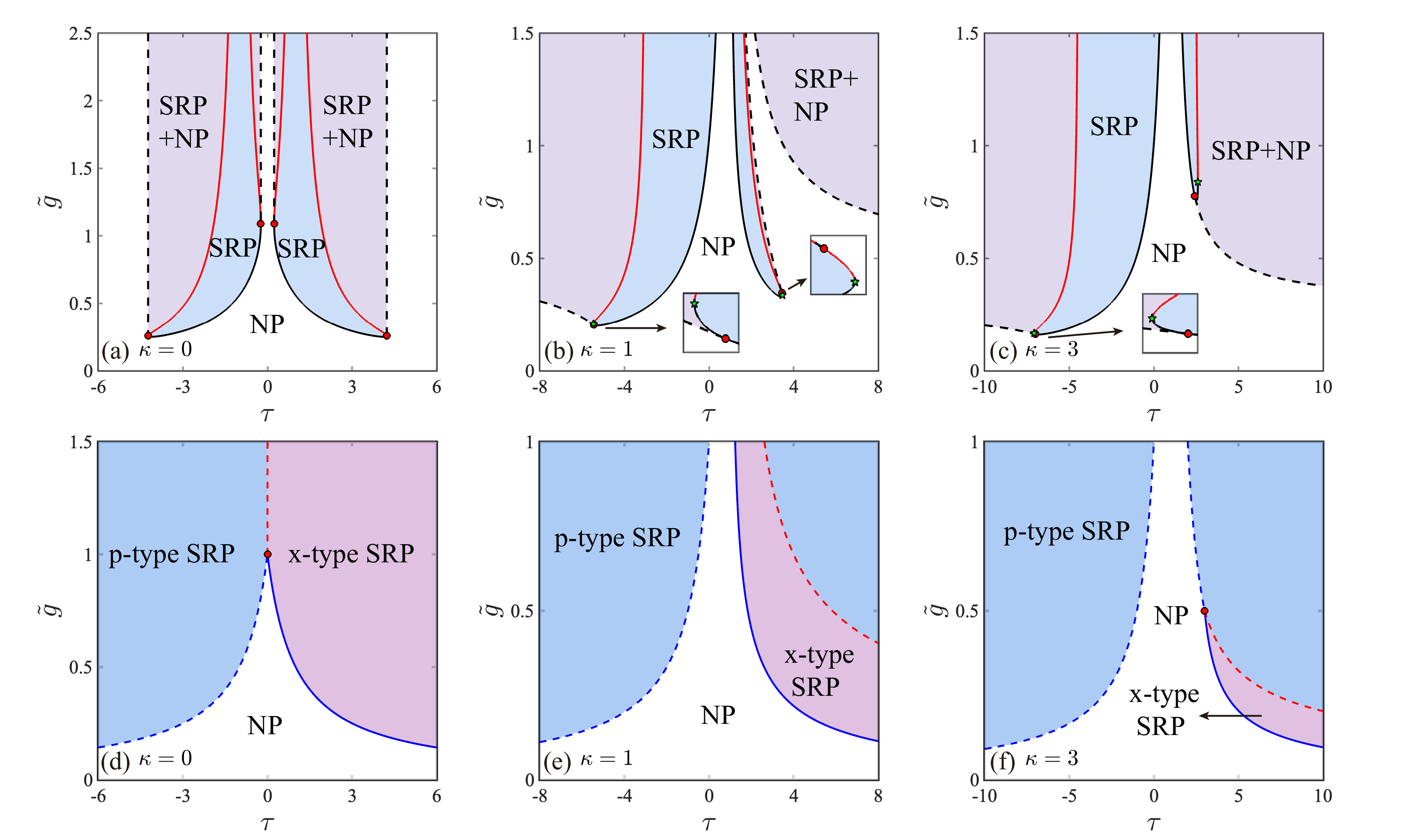}
\caption{Upper panels (a)--(c) present  the steady-state phase diagrams of the dissipative anisotropic QRM in the $\tilde{g}-\tau$ plane for $\kappa = 0$, $1$, and $3$, respectively. Regions corresponding to the NP (white), SRP (blue), and NP+SRP (purple) are marked accordingly. The critical couplings $\tilde{g}_c^{-}$ and $\tilde{g}_c^{+}$ are depicted as solid black and red curves, respectively. A tricritical point is indicated by a red circle, and the intersection of $\tilde{g}_c^{+}$ and $\tilde{g}_c^{-}$ is marked by a green pentagram. The boundary between the NP+SRP and NP phases is shown as a black dashed curve, with the dissipation rate fixed at $\gamma = 0.5\omega$.  Lower panels (d)--(f) display  the ground-state phase diagrams of the dissipationless  anisotropic  QRM  for the same $\kappa$ values as in the upper row. Here, dark-purple and dark-blue regions correspond to the $x$-type and $p$-type SRPs, respectively. The critical coupling strengths $\tilde{g}_c^{x}$ and $\tilde{g}_c^{p}$ are shown as solid and dashed blue lines. A first-order phase boundary between the two SRPs is plotted as a red dashed line, with red circles marking the tricritical points.
}
\label{fig:phase_diagram}
\end{figure*}

\subsubsection{General dissipative phase diagram}
Throughout this work, we focus on the thermodynamic limit $\Delta/\omega\to\infty$, where the DPT is sharply defined and the
mean-field treatment becomes exact, allowing the steady-state phase diagram to be determined accurately. Effects associated with finite
values of $\Delta/\omega$, which lead to finite-size rounding, are outside the scope of the present phase diagram analysis. The mean-field steady-state solutions are obtained by solving the
semiclassical equations of motion. This approach accurately yields the phase diagram in the limit of $\Delta /\omega \rightarrow \infty$. By neglecting
quantum fluctuations and factorizing operator expectation values, the
mean-field equations can be derived from Eq.~\eqref{Hamiltonian1} within the
Lindblad formalism given by Eq.~\eqref{lindblad}. For clarity, we introduce the renormalized parameters, leading to the following steady-state equations:
\begin{equation}
\begin{aligned} \left(1 - i\tilde{\gamma}\right)\alpha +
\tilde{g}\left(s_-+\tau s_+\right) + 2\kappa\tilde{g}^2\left(\alpha
+\alpha^{*}\right) &=0,\\ \tilde{g}\left(\alpha + \tau \alpha^*\right)s_z -
s_- &=0,\\ \alpha^*\left(s_- \tau -s_+\right) +\alpha\left(\tau s_-
-s_+\right) &=0, \label{steady} \end{aligned}
\end{equation}%
where the dimensionless quantities are defined as $\alpha =\sqrt{\omega
/\Delta }\langle a\rangle $, $\tilde{g}=g/\sqrt{\omega \Delta }$, and $%
\tilde{\gamma}=\gamma /\omega $. The spin mean values are denoted by $%
s_{x,y,z}=\langle \sigma _{x,y,z}\rangle $, and $s_{+}=(s_{x}+is_{y})/2$.
The mean displacement $\alpha $ is further decomposed as $\alpha =x+iy$.

Under these definitions, the steady-state equation Eq.~\eqref{steady} can be
rewritten as a set of four linear equations, $\mathcal{M}(x,y,s_{x},s_{y})^{%
\mathrm{T}}=0$. The trivial steady-state solutions, characterized by $%
x=y=s_{x}=s_{y}=0$ and $s_{z}=\pm 1$, preserve the $\mathbb{Z}_{2}$-symmetry. When the determinant of the matrix $\mathcal{M}$ vanishes, spontaneous symmetry breaking
occurs, signaling the onset of the SRP with the nonzero mean-field solutions. The corresponding nontrivial expressions for
$s_{z}$ are then given by
\begin{equation}
s_{z}=-\frac{h- \sqrt{h^2-q}}{\left( 1-\tau ^{2}\right) ^{2}\tilde{g}^{2}},
\label{sz}
\end{equation}%
where $h=1+\tau ^{2}+2\kappa \tilde{g}^{2}(1-\tau )^{2}$ and $q=(1-\tau ^{2})^{2}(1+4\kappa \tilde{g}^{2}+\tilde{\gamma}^{2})$. Thus, the semiclassical steady-state solution of the open anisotropic QRM exhibits a bifurcation from a zero mean-field solution to a symmetry-breaking  superradiant mean-field solution.

Finally, imposing the
spin conservation $s_{x}^{2}+s_{y}^{2}+s_{z}^{2}=1$ together with the
relations $s_{x}=2\tilde{g}(1+\tau )xs_{z}$ and $s_{y}=2\tilde{g}(\tau
-1)ys_{z}$, the nontrivial steady-state solutions for $x$ and $y$ are
obtained (see Appendix.~\ref{appendix A} for details). Furthermore, the
condition $s_z$ must be a real number requires $h^2\geq q$. The
boundary is given by $h^2=q$. For  $\kappa \neq 0$,
\begin{equation}
\tilde{g}_{c}^{b}=\sqrt{\frac{2\tau \pm \tilde{\gamma}\left( 1-\tau
^{2}\right) }{2\kappa \left( \tau -1\right) ^{2}}}.  \label{gcb}
\end{equation}%
For $\kappa =0,$ we obtain
\begin{equation}
\tau _{c}^{b}=\pm \frac{1\pm \sqrt{\tilde{\gamma}^{2}+1}}{\tilde{\gamma}},
\label{taob}
\end{equation}%
which corresponds to four solutions, all independent of the coupling strength $\tilde{g}$.

 The critical coupling strength $\tilde{g}_{c}$ between the NP and the SRP is
obtained  by setting $s_{z}=-1$, yielding
\begin{equation}
\tilde{g}_{c}^{\pm }=\sqrt{\frac{\tau ^{2}-2\kappa +1\pm \sqrt{4(\kappa
-\tau )^{2}-\tilde{\gamma}^{2}\eta }}{\eta }},  \label{g_c_np}
\end{equation}%
where $\eta =(\tau -1)^{2}[(1+\tau )^{2}-4\kappa ]$. The two branches $\tilde{g}_{c}^{+}$ and  $\tilde{g}_{c}^{-}$ 
coincide when  $4(\kappa -\tau )^{2}=\tilde{\gamma}^{2}\eta $,  which can be rewritten as
\begin{equation}
\left\vert \frac{\kappa -\tau }{1-\tau }\right\vert =\tilde{\gamma}\sqrt{{\left(
\frac{1+\tau }{2}\right)^{2}-\kappa }}. \label{tao1}
\end{equation}
{Thus, for a fixed $\kappa$, there exist  critical values $\tau_c^s$ at which $\tilde{g}_{c}^{\pm}$ merges.}

Now we construct the complete steady-state phase diagram of the open anisotropic QRM, as shown in the upper panels of Fig.~\ref{fig:phase_diagram} for $\kappa=0$ \hyperref[fig:phase_diagram]{(a)}, $\kappa=1$ \hyperref[fig:phase_diagram]{(b)}, and $\kappa=3$ \hyperref[fig:phase_diagram]{(c)} in the $(\tau,\tilde g)$ plane. For comparison, we reproduce and extend the phase diagram for $\kappa=0$ originally reported in Ref.~\cite{lyu2024PERL}. The NP--SRP boundaries, given by $\tilde g_c^\pm$ in Eq.~\eqref{g_c_np}, are shown as solid red and black curves, respectively. These two boundaries can equivalently be obtained from the linear stability condition of the NP, and therefore mark where the normal fixed point loses or recovers stability. The boundary $\tilde g_c^b$, determined by the reality condition $h^2=q$ for the nonzero mean-field solutions, gives the stability boundary of the SRP.

The region enclosed between the $\tilde{g}_{c}^{b}$ curve and the $\tilde{g}_{c}^{\pm}$ curves corresponds to the NP+SRP phase. In this region, both the NP and SRP are linearly stable, and the mean-field steady-state is selected by the initial state. This mean-field bistability relies on taking the frequency limit $\Delta/\omega\rightarrow\infty$ before the long-time limit. At finite $\Delta/\omega$, the exact Lindblad dynamics may instead yield a unique stationary state, with the NP+SRP coexistence appearing as metastability. For $\kappa \neq 0$, an isolated NP+SRP region—bounded only by the $\tilde{g}_{c}^{b}$ curve—also emerges, as shown in \hyperref[fig:phase_diagram]{(b)}. Such an isolated region can also appear in the regimes of large $\tilde{g}$ and $\tau$ in \hyperref[fig:phase_diagram]{(c)} (though it lies outside the range displayed in this figure), due to the two solutions of $\tilde{g}_{c}^{b}$ in Eq. \eqref{gcb}. In contrast to the case without the $A^2$ term in \hyperref[fig:phase_diagram]{(a)}, where the phase diagram is symmetric in $\tau$, the inclusion of the $A^2$ term breaks this symmetry and results in an asymmetric phase structure.

Importantly, as shown in Fig.~\ref{fig:phase_diagram}(b) and \hyperref[fig:phase_diagram]{(c)} for $\kappa \geqslant 1$, no superradiant phase transition occurs along the line $\tau = 1$. {In this sense, the "no-go theorem" established for closed systems~\cite{A2original,keeling2007jpb} extends to the open quantum Rabi model as well. Nevertheless, when $\tau \neq 1$, the superradiant phase transition does take place. Crucially, this anisotropy enables the system to bypass the limitation.}

Tricritical points, marked by red circles, arise from the intersection of the NP, SRP, and NP+SRP phases. They correspond to the meeting points of the NP stability boundary $\tilde g_c^\pm$ and the SRP stability boundary $\tilde g_c^b$ in mean-field framework. Each such point corresponds to the crossing of the $\tilde{g}_{c}^{b}$ curve and one branch of the $\tilde{g}_{c}^{\pm}$ transition lines. Notably, once the $A^2$ term is present, this tricritical point no longer coincides with the intersection point $\tilde{g}_{c0}$ (green pentagram)  of the $\tilde{g}_{c}^{\pm}$ lines, as illustrated in the insets of Fig.~\ref{fig:phase_diagram} (b) and \hyperref[fig:phase_diagram]{(c)}. 

This distinction can be directly verified: for fixed  $\tilde{\gamma}$ and nonzero $\kappa$, the value of $\tau$ determined from Eq.~\eqref{tao1} cannot simultaneously satisfy $\tilde{g}_{c}^{b} = \tilde{g}_{c}^{+} = \tilde{g}_{c}^{-}$. By contrast, in the case of $\kappa = 0$, $\tilde{g}_{c}^{b}$ becomes arbitrary. Hence, the tricritical point is determined solely by the intersection of the $\tilde{g}_{c}^{\pm}$ lines~\cite{lyu2024PERL}. In contrast to the $\kappa=0$ case where the tricritical point is defined by the $\tilde{g}_{c}^{\pm}$ intersection, its position changes once the $A^2$ term is included, as shown in Fig.~\ref{fig:phase_diagram}(b, c). This relocation results in a modified critical behavior near the tricritical point, which will be examined in Sec.~\ref{sec4}.

\subsubsection{Special Cases and Previous Limits}
It is instructive to verify that the above nontrivial solutions reproduce known results in the absence of the $A^2$ term or in the dissipationless limit ($\tilde{\gamma} \to 0$). 

First, upon setting $\kappa = 0$, the expressions for $s_z$ and $\tilde{g}_c^{\pm}$ reduce to those obtained by Lyu et al. \cite{lyu2024PERL}:
\begin{equation}
\tilde{g}_{c}^{\pm} = \frac{\sqrt{\tau ^{2} + 1 \pm \sqrt{4\tau ^{2} - \tilde{\gamma}^{2}\left( 1 - \tau ^{2}\right)^{2}}}}{\left\vert 1 - \tau ^{2}\right\vert }.
\end{equation}
In the symmetric QRM and in the absence of the $A^2$ term ($\tau = 1, \kappa = 0$), the solutions simplify to $s_{z}=-{(1+\tilde{\gamma}^2)}/{4\tilde{g}^{2}}
$, $\alpha =\pm \frac{\tilde{g}}{1-i\tilde{\gamma}}\sqrt{1-(1+\tilde{\gamma}%
^{2})^{2}/\left( 16\tilde{g}^{4}\right) }$, and the critical coupling
strength becomes $g_{c}^{-}=\sqrt{1+\tilde{\gamma}^{2}}/{2}$, which
reproduces previously established results \cite{Hwang2018PRA}. On the other hand, when the $
A^2$ term is included and the TRK condition $\kappa >1$ is
imposed in the symmetric case, the determinant of the coefficient matrix
becomes:
\begin{equation}
\det [\mathcal{M}]=-\frac{1}{4}\!\left[ (1+4\kappa \tilde{g}^{2}+\tilde{%
\gamma}^{2})+4\tilde{g}^{2}s_{z}\right] .
\end{equation}%
Setting $\det [\mathcal{M}]=0$ yields the unphysical solution $s_{z}=-(1+%
\tilde{\gamma}^{2}+4\kappa \tilde{g}^{2})/(4\tilde{g}^{2})<-1$. This implies
that the DPT is prohibited in the symmetric open QRM.

Second, taking the dissipationless limit ($\tilde{\gamma} \to 0$) corresponding to a closed system, Eq.~\eqref{g_c_np} reduces to
\begin{equation}
\tilde{g}_{c}^{\pm }=\sqrt{\frac{\tau ^{2}-2\kappa +1\pm \sqrt{4(\kappa
-\tau )^{2}}}{(\tau -1)^{2}[(1+\tau )^{2}-4\kappa ]}},
\end{equation}%
which yields
\begin{equation}
\begin{aligned} \tilde{g}_{c}^{-} &= \frac{1}{\left| 1-\tau \right|}, \quad
&\tau < \kappa, \\ \tilde{g}_{c}^{-} &= \frac{1}{\sqrt{(\tau+1)^2 -
4\kappa}}. \quad &\tau > \kappa, \end{aligned} \label{closed}
\end{equation}

{These expressions coincide with the critical couplings $\tilde{g}_{c}^{p}$ and $\tilde{g}_{c}^{x}$ of the closed model, i.e., the critical strengths for the p-type and x-type SRPs which  are clearly distinguished by their real and purely imaginary coherent amplitudes, respectively~\cite{ye2025PRA}.} Notably, both originate from the branch $\tilde{g}_{c}^{-}$, while $\tilde{g}_{c}^{+}$ has no counterpart  in the dissipationless (closed) system.

To establish a direct connection with the dissipationless limit, the lower panels of Fig.~\ref{fig:phase_diagram} reproduce the ground-state phase diagrams of the closed anisotropic QRM from Ref.~\cite{ye2025PRA} at corresponding $\kappa$ values. Consequently, the black solid line in Fig.~\ref{fig:phase_diagram}(c) traces the same phase boundary as the ensemble of blue dashed and solid lines in panel (f). Similarly, the first-order phase transition boundary for the ground state, obtained from Eq.~\eqref{gcb} with $\tilde{\gamma} = 0$ as $\tilde{g}_{c}^{f} = \sqrt{\tau /[\kappa (\tau - 1)^{2}]}$, coincides with the boundary separating the p-type and x-type SRPs; hence, the black dashed line in Fig.~\ref{fig:phase_diagram}(c) maps directly onto the red dashed line in panel \hyperref[fig:phase_diagram]{(f)}.

The two critical lines given by Eq. \eqref{closed} intersect at $\kappa = \tau > 1$, which defines the tricritical point $\tilde{g}_{c}^{\mathrm{tri}} = 1/(\kappa - 1)$. This expression can also be derived from Eq. \eqref{gcb} in the dissipationless limit ($\tilde{\gamma} \to 0$) and matches the closed-model result~\cite{ye2025PRA}. Therefore, in the right panels ($\kappa = 3$), both the tricritical point and the $\tilde{g}_{c}^{\pm}$ intersection point $\tilde{g}_{c0}$ continuously evolve and coalesce into the ground-state tricritical point (red circle in Fig.~\ref{fig:phase_diagram}(f) as $\tilde{\gamma} \to 0$). Moreover, the tricritical points at $\tau < \kappa$ have no counterpart in the closed system. 

As $\tilde{\gamma}\rightarrow 0$, one branch of Eq.~\eqref{taob} for $\kappa=0$ reduces to
$\tau _{c}^{b}=\mp \tilde{\gamma}/2\rightarrow 0$. Furthermore, from Eq. ({\ref{g_c_np}}), we obtain $\tilde{g}_{c}^{\pm }=1/(1\mp \tau)$. Thus, in the absence of the $A^2$ term, the tricritical points approach $\tau = 0$ and $\tilde{g} = 1$ in the dissipationless limit, which coincides with the exact tricritical point in the closed system, as illustrated  in Fig.~\ref{fig:phase_diagram}(d). By contrast, the tricritical points appearing in Fig.~\ref{fig:phase_diagram}(b) have no counterpart in the closed system, which itself exhibits no tricritical points. Indeed, this holds for all cases where $0 < \kappa \leq 1$.

Interestingly, in the open anisotropic QRM, the emergence of the new NP+SRP phase results in tricritical points for both positive and negative $\tau$—in sharp contrast to the closed QRM, where  tricritical points exist only for $\tau > 0$. Notably, the global phase diagram in Fig.~\ref{fig:phase_diagram}(a) contains four such tricritical points, all situated on the four vertical lines specified by Eq.~\eqref{taob} for $\kappa = 0$.

To facilitate direct comparison with known results for the open anisotropic QRM (e.g., Ref.~\cite{lyu2024PERL}), we present the same steady-state phase structure as in the upper panels of Fig.~\ref{fig:phase_diagram} on the positive coupling $(g_{cr}=\tau \tilde{g}>0, g_{r}=\tilde{g}>0)$ plane in Fig.~\ref{fig:phase_diagram1} of Appendix~\ref{appendix A}. The latter is therefore a complementary representation of Fig.~\ref{fig:phase_diagram}, introduced to compare directly with the conventional $(g_{cr},g_r)$ notation. For reference, the phase diagram without the $A^2$ term and that of the closed QRM are replotted in the left panels. For $\kappa\geq1$, the isotropic line $g_{cr}=g_r$ stays in the NP, demonstrating that the DPT is forbidden by the $A^2$ term, as shown in Fig.~\ref{fig:phase_diagram1}(b,c).

\subsection{Stability analysis}

To examine the dynamical stability of the mean-field solutions, we analyze the
stability of both the NP and the SRP. Small fluctuations around the mean-field
steady states are introduced as
$\alpha \rightarrow \alpha + \delta\alpha$,
$s_- \rightarrow s_- + \delta s_-$, and
$s_z \rightarrow s_z + \delta s_z$. Retaining only terms linear in these fluctuations yields a set of
linearized equations of motion around the mean-field solutions, which
capture the dynamics of small perturbations and allow for an assessment of
the dynamical stability of both the NP and the SRP.

In the limit $\Delta/\omega \rightarrow \infty$, the spin dynamics becomes
much faster than the cavity dynamics and adiabatically follows the cavity
field. Accordingly, the spin fluctuations can be eliminated by imposing $%
d\delta s_x/dt = 0$ and $d\delta s_y/dt = 0$, corresponding to their
instantaneous steady-state conditions. Within this adiabatic approximation,
the spin fluctuations are expressed in terms of the cavity-field
fluctuations as
{\small
\begin{equation}
\begin{aligned} \delta s_x &=\frac{2\tilde{g} s_z \left(1 +
\tau\right)}{\Sigma} \bigg\{\left[4\tilde{g}^2\left(1-\tau\right)^2y^2 +
1\right]\delta x- 4\tilde{g}^2\left(1-\tau\right)^2x y \delta y
\bigg\},\\ \delta s_y &=\frac{2\tilde{g} s_z \left(1 - \tau\right)}{\Sigma}
\bigg\{4\tilde{g}^2\left(1+\tau\right)^2x y \delta x-\left[4\tilde{g}^2\left(1+\tau\right)^2x^2 + 1\right]\delta y\bigg\}.
\end{aligned}  \label{adiabation}
\end{equation}
}
where $\Sigma = 1 + 4\tilde{g}^2\!\left[\left(1+\tau\right)^2 x^2 +
\left(1-\tau\right)^2 y^2\right]$. As a result, the spin degrees of freedom
can be adiabatically eliminated, yielding an effective stability matrix $%
\mathbb{M}$ that depends solely on the cavity-field operators [see Appendix.~\ref{appendix B} for details].

The eigenvalues of the effective stability matrix $\mathbb{M}$ are $\lambda_{\pm}=\omega[-\tilde{\gamma}\pm\sqrt{\tilde{\gamma}^{2}-Q/\Sigma}]$. A mean-field solution is dynamically stable when $\mathrm{Re}\lambda_{\pm}<0$. Since $\Sigma>0$, this stability condition is equivalent to $Q>0$, where $Q=(1+\tilde{\gamma}^{2}+4\kappa\tilde{g}^{2})\Sigma+2(1+\tau^{2})s_z\tilde{g}^{2}+(1-\tau^{2})^{2}[s_z^{2}+4(x^{2}+y^{2})s_z]\tilde{g}^{4}+4\kappa s_z(1-\tau)^{2}\tilde{g}^{4}[4\tilde{g}^{2}(1+\tau)^{2}x^{2}+1]$. For the NP, where $x=y=0$ and $s_z=-1$, this expression reduces to $Q_{\mathrm{NP}}=1+\tilde{\gamma}^{2}+2(2\kappa-\tau^{2}-1)\tilde{g}^{2}+(1-\tau)^{2}[(1+\tau)^{2}-4\kappa]\tilde{g}^{4}$. Therefore, the NP is stable for $Q_{\mathrm{NP}}>0$, while $Q_{\mathrm{NP}}=0$ gives the NP stability boundaries $\tilde{g}_{c}^{\pm}$ in Eq.~\eqref{g_c_np}, corresponding to the black and red solid curves in Fig.~\ref{fig:phase_diagram}. For the SRP, the same criterion $Q>0$ applies after substituting the nonzero mean-field solutions into $Q$. In bistable phase, both the NP and SRP satisfy the stability condition.

 Interestingly, a bistable region emerges where the NP and
SRP coexist. Crossing the critical line $\tilde{g}_c^{\pm}$   triggers a second-order dissipative phase transition between NP and SRP. Notably, even
within the NP region, the onset of the SRP begins at the boundary
$\tilde{g}_c^b$, where $s_{z}$ becomes real, yielding a finite $\alpha$,
consequently, a stable SRP also emerges.

\begin{figure}
\centering
\includegraphics[width=0.45\textwidth]{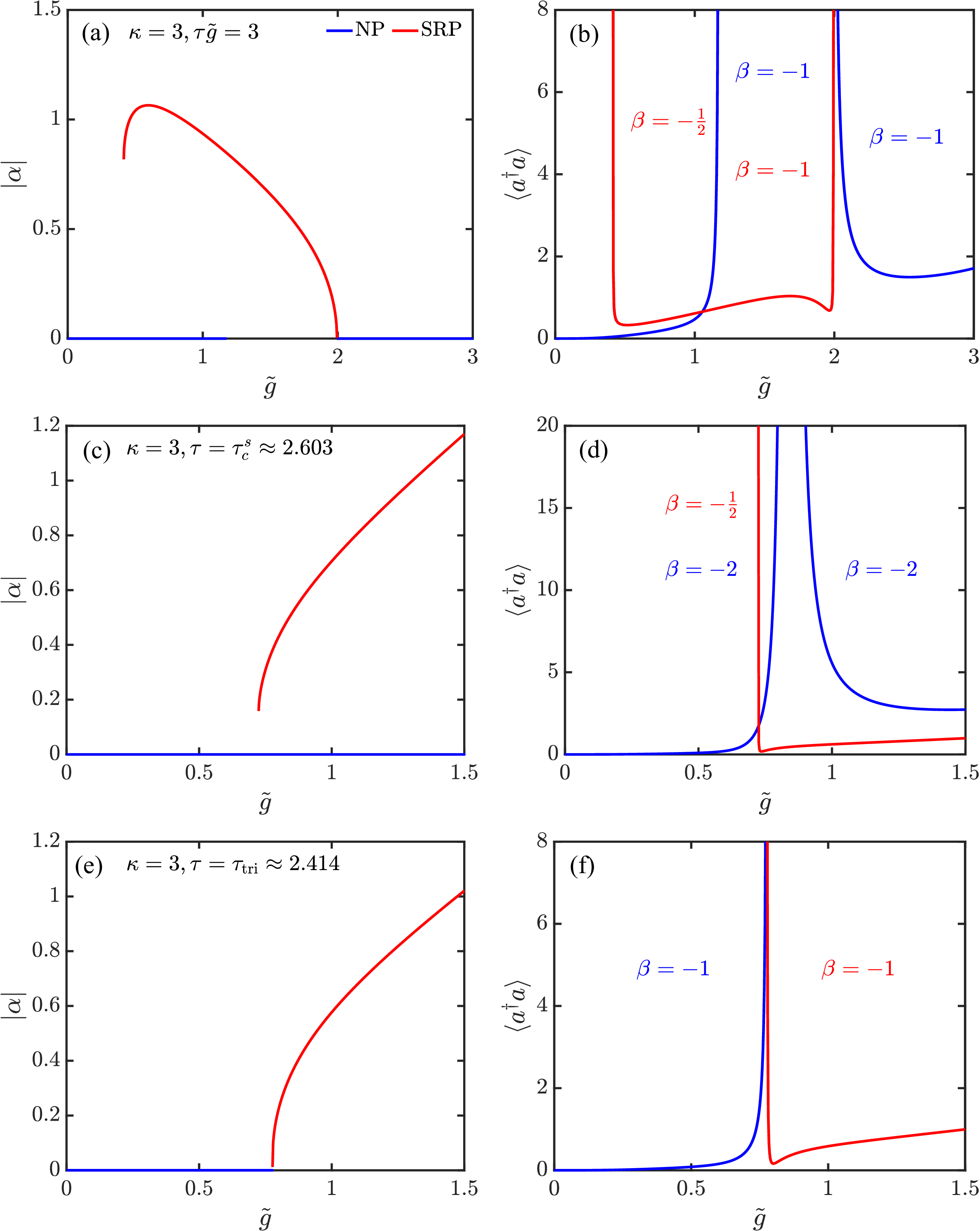}
\caption{{The left panels display the order parameter $|\alpha|$ along three representative parameter paths:
(a) a path with fixed CRW coupling strength $g_{cr}=\tau \tilde{g} = 3$, corresponding to the black dotted line in Fig.~\ref{fig:phase_diagram1}(c);
(c) a path  that crosses  the intersection point $\tilde{g}_{c0}$ of $\tilde{g}_c^{\pm}$ lines at fixed $\tau = \tau_{c}^s \approx 2.603$;
and (e) a path passing through the tricritical point at fixed $\tau = \tau_{\mathrm{tri}} \approx 2.414$.
The right panels show the photon-number fluctuations along the same parameter paths, with
(b), (d), and (f) corresponding to panels (a), (c), and (e), respectively. Solid blue and red curves represent the NP and the SRP, respectively. Other parameters are fixed at $\kappa = 3$ and $\gamma = 0.5\omega$.} }
\label{fig:alpha_fluctuation}
\end{figure}

The order parameter $|\alpha|$ is shown in Fig.~\ref{fig:alpha_fluctuation}(a) to characterize the nature of the DPTs, with particular attention paid to the transition between the bistable phase and the NP. To this end, we fix the dimensionless CRW coupling strength at
$g_{cr}=\tau \tilde{g} = 3$ for $\kappa = 3$, as indicated by the black dotted line in
Fig.~\ref{fig:phase_diagram1}(c).
As $\tilde{g}$ increases, the system undergoes successive phase transitions: NP  $\rightarrow$ NP+SRP $\rightarrow$ SRP$\rightarrow$ NP.

In panel \hyperref[fig:alpha_fluctuation]{(a)}, along the $g_{cr} =\tau \tilde{g} = 3$ line, as $\tilde{g}$ crosses $\tilde{g}_c^b\approx 0.415$, the order parameter $|\alpha|$ exhibits a discontinuous jump from zero to a finite value, marking a first-order DPT from the NP to the bistable phase (NP+SRP). Further increasing the coupling beyond $\tilde{g}_{c}^{+} \approx 1.173$ causes the NP to lose dynamical stability, driving the system into the SRP. When $\tilde{g}$ exceeds $\tilde{g}_{c}^{-} \approx 1.995$, $|\alpha|$ continuously decays to zero, indicating a second-order DPT back to the NP. These results underscore the rich interplay between dissipation and anisotropy in governing the DPTs.

As shown in Fig.~\ref{fig:alpha_fluctuation}(c), we fix $\tau = \tau_c^s \approx 2.603$ along a vertical line passing through the intersection point of $\tilde{g}_{c0}=\tilde{g}_c^{\pm}$ in Fig.~\ref{fig:phase_diagram} (c). 
Along this path, the system undergoes a first-order dynamical DPT from the NP to NP+SRP, characterized by a sudden jump of $|\alpha|$ when $\tilde{g}$ exceeds $\tilde{g}_c^b \approx 0.724$. Meanwhile, the NP remains stable throughout, and no phase transition occurs at $\tau = \tau_c^s$. However, the fluctuations of the mean photon number exhibit a higher-order critical exponent at the intersection point $\tilde{g}_{c0}\approx0.839$, as discussed in Sec.~\ref{sec4}.

In Fig.~\ref{fig:alpha_fluctuation}(e), $|\alpha|$ is shown along the vertical line that passes through the tricritical point in Fig.~\ref{fig:phase_diagram} (c). Along this line, $|\alpha|$ becomes nonzero via a second-order DPT from the NP to the SRP. This behavior occurs because the $A^2$ term shifts the tricritical point, causing it to no longer coincide with the intersection of the $\tilde{g}_{c}^{\pm}$ boundaries. As a result, the system does not enter a bistable phase along this path.

\section{\label{sec4} Effective master equation and Quantum fluctuations}

In this section, effective master equations governing the full quantum
dynamics are derived in the frequency limit $\Delta/\omega \rightarrow \infty
$, where the resulting description is characterized by a quadratic form in
the cavity-field operators. The quantum fluctuations of the cavity-field are
then obtained from the equations of motion for the second moments of
cavity-field operators.

\subsection{Normal phase}

In the NP, the photon number in the cavity is close to zero and the
steady-state solution is given by $\alpha = 0$. We therefore perform the
following sequence of transformations to derive the effective master
equation. Firstly, a squeezing transformation $S^\dagger H S$ is applied
using the squeezing operator $S = \exp\left[\frac{r}{2}\left(a^{%
\dagger2}-a^2\right)\right]$ with $r = -\frac{1}{4}\ln\left(1+4\kappa \tilde{g}^2\right)$ to eliminate the $A^2$ term. Subsequently, a
Schrieffer--Wolff (SW) transformation is performed, $H_{\text{NP}} =
V^\dagger H^\prime V$, where $V = \frac{\tilde{g}}{\Delta}
\left(a^\dagger\sigma_{-}-a\sigma_{+}\right)-\frac{\tilde{g}\tau}{\Delta}
\left(a^\dagger \sigma_{+} - a \sigma_{-}\right)$. Retaining terms up to
second order in the small parameter $\omega/\Delta$ and projecting onto the
spin-down subspace $\ket{\downarrow}$, the resulting low-energy effective
Hamiltonian is obtained as
\begin{equation}
H_{\text{NP}}^{\downarrow} = R_{\text{NP}}a^\dagger a + P_{\text{NP}
}a^{\dagger 2} + P_{\text{NP}}^* a^{2}  \label{H_np}.
\end{equation}
Here, $R_{\mathrm{NP}} = \frac{\omega}{2}\!\left[ 2\xi - \tilde{g}^2 \left(
\frac{(1+\tau)^2}{\xi} + \xi (1-\tau)^2 \right) \right]$ and
$P_{\mathrm{NP}} = \frac{\omega}{4}\,\tilde{g}^2 \!\left[ \xi (1-\tau)^2 -
\frac{(1+\tau)^2}{\xi} \right]$,
where $\xi = \sqrt{1 + 4\kappa \tilde{g}^2}$. The corresponding effective master equation for the NP is obtained by applying
the same sequence of transformations to the original Lindblad equation~%
\eqref{lindblad}, yielding $\dot{\rho}_{\text{NP}} = -i\left[H_{\text{NP}%
}^{\downarrow},\rho_\text{NP}\right] + \gamma \mathcal{D}\left[\tilde{a}%
\right]$, where $\tilde{a} = V^\dagger S^\dagger a S V = a\cosh r +
a^\dagger \sinh r$, $\rho_{\text{NP}} = \bra{\downarrow} V^\dagger S^\dagger
\rho S V \ket{\downarrow}$ and the effective dissipative super-operator $%
\mathcal{D}[\tilde{a}] = 2\tilde{a}\rho_{\text{NP}} \tilde{a}^\dagger -
\tilde{a}^\dagger \tilde{a} \rho_{\text{NP}} - \rho_{\text{NP}} \tilde{a}%
^\dagger \tilde{a}$. Only the first moments of the field operators are
retained. The resulting equations of motion for $\mathbf{a}=\left(\braket{a}%
, \braket{a^\dagger}\right)^T$ are derived from the effective Hamiltonian~%
\eqref{H_np} as
\begin{equation}
\dot{\mathbf{a}} = L_{\text{NP}} \mathbf{a},
\end{equation}
with the Liouville matrix
\begin{equation}
L_{\text{NP}}=%
\begin{pmatrix}
-iR_{\text{NP}} - \gamma  & -2iP_{\text{NP}} \\
2iP_{\text{NP}}^* & iR_{\text{NP}}-\gamma
\end{pmatrix}%
.  \label{Liouville_np}
\end{equation}
The eigenvalues $l_{\text{NP}}^{\pm}$ of $L_{\text{NP}}$ are given by $l_{%
\text{NP}}^{\pm} = -\gamma \pm \sqrt{ 4 |P_{\text{NP}}|^2 - R_{\text{NP}}^2}$. The NP is dynamically
stable when the real parts of all eigenvalues are negative. It is noted that
$l_{\text{NP}}^{-}$ is always negative. The condition $l_{\text{NP}}^{+}=0$
also yields stability boundaries that are consistent with the critical
couplings $\tilde{g}_c^{\pm}$ given in Eq.~\eqref{g_c_np}.

\subsection{Superradiant phase}

Photons in the cavity is macroscopically populated when the system enters
the SRP. Firstly, a displacement transformation is applied to the
Hamiltonian~\eqref{Hamiltonian1} using the displacement operator  $%
D\left(\alpha\right)=\exp\!\left[\sqrt{\Delta/\omega}\left(\alpha
a^\dagger-\alpha^* a\right)\right]$. And the transformed master equation is
described by $\dot{\bar{\rho}} = -i\left[\bar{H},\bar{\rho}\right] + \gamma
\mathcal{D}\left[{a}\right]$, where $\bar{\rho} = D^\dagger \rho D$, $\bar{H}
= D^\dagger H D + i\sqrt{\Delta \omega}\tilde{\gamma}\left(\alpha^* a
-\alpha a^\dagger\right)$. The second term of $\bar{H}$ describe only spin
part and it can be diagonalized as $U^\dagger \bar{H}_s U = \frac{\Delta}{2}
\sqrt{1+4\tilde{g}^2|\alpha + \tau \alpha^*|^2}\sigma_z = \frac{%
\Delta\sigma_z}{2|s_z|}$ by the unitary operator
\begin{equation}
U = \frac{1}{\sqrt{2\chi+2\sqrt{\chi}}}%
\begin{pmatrix}
1 + \sqrt{\chi} & -2\tilde{g}\left(\alpha^* + \tau \alpha\right) \\
2\tilde{g}\left(\alpha + \tau \alpha^*\right) & 1+\sqrt{\chi}%
\end{pmatrix}%
,
\end{equation}
where $\chi = 1/s_z^2$. And $\bar{H}^{\prime}=U^\dagger \bar{H} U$ is
transformed Hamiltonian in new spin basis. The linear terms of $\bar{H}%
^{\prime}$ vanish after projection onto the spin-down subspace $%
\ket{\downarrow}$. Subsequently, the same squeezing transformation $\bar{H}%
^{\prime\prime} = S^\dagger \bar{H}^\prime S$ is applied to eliminate the $%
A^2$ term.  Finally, a SW transformation $H_{\text{SRP}} =
V^\dagger \bar{H}^{\prime\prime} V$ is performed, followed by projection
onto the spin-down subspace $\ket{\downarrow}$, yielding the effective
Hamiltonian in the SRP, which retains the same mathematical structure as NP,
\begin{equation}
\begin{aligned} H_{\text{SRP}}^{\downarrow} &= R_{\text{SRP}}a^\dagger a +
P_{\text{SRP}}a^{\dagger 2} + P_{\text{SRP}}^* a^{2}. \end{aligned}
\label{H_srp}
\end{equation}
The expressions for the coefficients $R_{\text{SRP}}$ and $P_{\text{SRP}}$
are lengthy and are provided in Appendix~\ref{appendix C}. The corresponding
effective master equation for the SRP takes the form $\dot{\rho}_{\text{SRP}%
}=-i\left[H_{\text{SRP}}^{\downarrow},\rho_{\text{SRP}}\right] + \gamma
\mathcal{D}\left[\bar{a}\right]$, where $\rho_{\text{SRP}}= \bra{\downarrow}
V^\dagger S^\dagger U^\dagger \bar{\rho} U S V\ket{\downarrow}$ and $\tilde{a%
} = V^\dagger S^\dagger U^\dagger a U S V$ [see Appendix.~\ref{appendix C}
for details]. The Liouville matrix retains the same mathematical structure
as Eq.~\eqref{Liouville_np} and the SRP is dynamically stable when the real
parts of all eigenvalues $l_{\text{SRP}}^{\pm}$ are negative.

\subsection{Quantum fluctuations and universality}

    \begin{figure*}
		\centering	\includegraphics[width=0.9\textwidth]{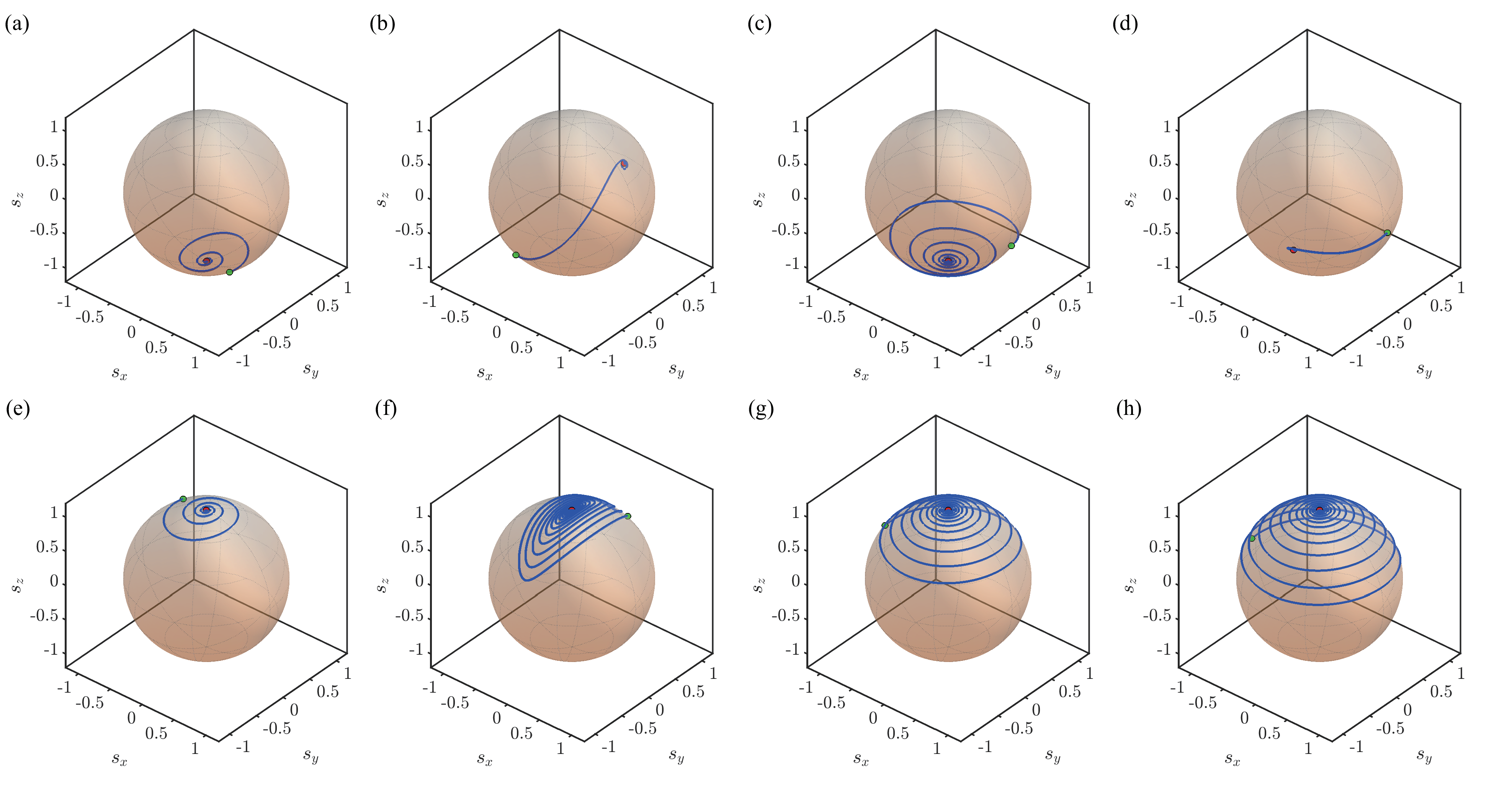}
		\caption{Bloch-sphere trajectories of the spin dynamics for different coupling strengths $\tilde{g}$ and anisotropy parameters $\tau$. Upper panels correspond to a distinct parameter set: (a) the NP at $(\tau = 0.5, \tilde{g} = 1)$ with the initial state $\alpha_0 = 0.2 + 0.2i$, (b) the SRP at $(\tau = 2, \tilde{g} = 1)$ with the initial state $\alpha_0 = 0.2 + 0.2i$, and (c–d) the bistable phase at $(\tau = 6, \tilde{g} = 0.5)$ with initial states $\alpha_0 = 0.3 + 0.05i$ (c) and $\alpha_0 = 0.7 + 0.1i$ (d). The trajectories evolve from the initial coherent states (green dots) toward their corresponding steady states (red dots), with the dynamical evolution represented by the blue curves. Lower panels (e–h), all trajectories start from initial conditions with $s_z > 0$ and converge to the spin-inverted steady state $s_z = 1$, with eachpanel corresponding to the same parameter settings as in the upper panels (a–d), respectively. Other system parameters are fixed at $\kappa = 3$ and $\gamma = 0.5\omega$.}
	\label{fig:spin_dynamics}
	\end{figure*}
    
Within the effective master equation framework, the quantum fluctuations of
the cavity field are investigated based on the effective Hamiltonians of the
NP (Eq.~\eqref{H_np}) and the SRP (Eq.~\eqref{H_srp}). The fluctuations are
characterized by the second moments of the field operators, defined as $\mathbf{s} = \left(\langle a^\dagger a \rangle, \langle a^2 \rangle, \langle
a^{\dagger 2} \rangle \right)^T$. Thus their dynamics is governed by the
equation of motion
\begin{equation}
\dot{\mathbf{s}} = K \mathbf{s} + Y,
\end{equation}
where
\begin{equation}
\begin{aligned} K &= \begin{pmatrix} -2\gamma &2iP^* & -2iP\\ -4iP
&-2iR-2\gamma &0\\ 4iP^* &0 &2iR-2\gamma \end{pmatrix}, \end{aligned}
\end{equation}
and $Y = \big(2\gamma \sinh^2 r,\, -2iP - \gamma \sinh 2r,\, 2iP^* -
\gamma \sinh 2r\big)^T$. Here, $P = P_{\text{NP}/\text{SRP}}$ and $R
= R_{\text{NP}/\text{SRP}}$ denote the coefficients for the NP and the SRP, respectively. The steady-state solution is then given by $\mathbf{s}_{s} = -K^{-1}Y$, and the photon-number fluctuations in the steady state are
obtained as
\begin{equation}
	\braket{a^\dagger a} = \frac{\tilde{R}^2 \sinh^2 r + \tilde{P}\sinh 2r + 2|P|^2}{\gamma^2 + R^2 - 4|P|^2},
    \label{fluctuation}
\end{equation}
where $\tilde{R}^2=R^2+\gamma^2$ and $\tilde{P}=-\gamma\mathrm{Im}\left(P\right)-R\mathrm{Re}\left(P\right)$. Based on the solutions obtained above, we now examine the universal critical
behavior associated with the DPTs in the anisotropic QRM including the $A^2$
term. The fluctuations of the local mean photon number near the critical
points diverge as
\begin{equation}
\braket{a^\dagger a}\propto |\tilde{g}-\tilde{g}_{c}|^{\beta },
\end{equation}
where $\beta$ denotes the critical exponent. {Meanwhile, the Liouvillian gap $l_{\mathrm{np/srp}}^{+}$ closes at the transition point, with the asymptotic decay
rate scaling as
$\mathrm{Re}[l_{\mathrm{np/srp}}^{+}] \propto
\lvert \tilde{g}-\tilde{g}_{c} \rvert^{\nu}$.
The two critical exponents are related by $\beta = -\nu$.}

{For the NP, the divergence of the mean photon number in the vicinity of the
critical couplings $\tilde{g}_c^{\pm}$ is governed by the vanishing of the
denominator,
$Q_{\mathrm{NP}} \propto \gamma^2 + R_{\mathrm{NP}}^2 - 4 |P_{\mathrm{NP}}|^2
= \left(\tilde{g}^2 - \tilde{g}_c^{+\,2}\right)
  \left(\tilde{g}^2 - \tilde{g}_c^{-\,2}\right)$.
Accordingly, as $\tilde{g} \to \tilde{g}_c^{\pm}$, the mean photon number
diverges with the critical exponent $\beta = -1$ as shown in Fig.~\ref{fig:alpha_fluctuation}(b, f), indicating a second-order
DPT that belongs to the Dicke universality class.
At the intersection point $\tau = \tau_c^{s}$, given by
Eq.~\eqref{tao1}, the two critical branches $\tilde{g}_c^{+}$ and
$\tilde{g}_c^{-}$ merge into a single critical point
$\tilde{g}_{c0}$. As a consequence, the denominator of
$\braket{a^\dagger a}_{\mathrm{NP}}$ reduces to
$\left(\tilde{g}^2 - \tilde{g}_{c0}^2\right)^2$, giving rise to a higher-order
divergence characterized by the critical exponent $\beta = -2$, as shown in
Fig.~\ref{fig:alpha_fluctuation}(d).
At this intersection point, the two branches $\tilde{g}_c^{\pm}$ share a common
tangent.
}

In the panel Fig.~\ref{fig:alpha_fluctuation}(b,d), near the critical point of the first-order DPT at $\tilde{g}_{c}^{b}$, a
distinct scaling behavior with critical exponent $\beta =-1/2$ emerges, signaling a new universality class associated with the bistable phase. This
result is consistent with the first-order DPT between the NP and the NP+SRP
bistable phase reported in Ref.~\cite{lyu2024PERL}.

The difference manifests in the scaling of the mean photon number along the vertical line that passes through the tricritical point in Fig.~\ref{fig:phase_diagram}(c). In the absence of the $A^2$ term, the scaling exponent $%
\beta =-2$ along the similar line was reported in Ref.~\cite{lyu2024PERL}. This
anomalous scaling originates from the coincidence of the $\tilde{g}_{c}^{\pm
}$ intersection point $\tilde{g}_{c0}$   with the tricritical point.

By contrast, when the $A^2$ term is included, the photon-number fluctuations at the tricritical point revert to the scaling exponent $\beta =-1$, as shown in Fig.~\ref{fig:alpha_fluctuation}(f). This occurs because the tricritical point now meets only one of the $\tilde{g}_{c}^{\pm}$ critical lines and not the intersection $\tilde{g}_{c0}$. Consequently, only one of the two factors $\left(\tilde{g}^2 - \tilde{g}_c^{+\,2}\right)$ and $\left(\tilde{g}^2 - \tilde{g}_c^{-\,2}\right)$ becomes relevant in the expression for $Q_{\mathrm{NP}}$. This shift, which moves the tricritical point away from $\tilde{g}_{c0}$, demonstrates that the $A^2$ term plays a decisive role in governing the critical behavior. In fact, near the tricritical point, the critical behavior is path dependent rather than universal: approaching the point along different paths in parameter space leads to distinct scaling behaviors \cite{riedel1974effective}.

\section{\label{sec6}Numerical validation and characterization of the steady‑state phases}
In this section, we first examine the evolution from an arbitrary initial state to the corresponding steady state in each parameter region of a given phase, thereby confirming the results obtained in the previous section. We then use the Wigner function to characterize the different steady-state phases, which can serve as a reference for experimental detection.

    \subsection{Spin dynamics}
	In the limit $\Delta/\omega \rightarrow \infty$, the spin dynamics evolves on a much faster timescale than the cavity field and can thus be adiabatically eliminated from the mean-field equations of motion. This procedure yields the relations $s_- = \tilde{g}\left(\alpha + \tau \alpha^*\right)s_z$ and $s_z = \pm {1}/{\sqrt{1 + 4\tilde{g}^2 \left|\alpha + \tau\alpha^*\right|^2}}$. Substituting these expressions into the cavity equation leads to a closed semiclassical motion equation for the cavity field:
	\begin{equation}
		\begin{aligned}
			\frac{d \alpha}{dt} &= -i\omega\left(1-i\tilde{\gamma}\right)\alpha \mp i\frac{\omega\tilde{g}^2\left[\left(1 + \tau^2\right)\alpha + 2\tau \alpha^{*}\right]}{\sqrt{1 + 4\tilde{g}^2 \left|\alpha + \tau \alpha^*\right|^2}} \\
            &- i 2\omega \kappa \tilde{g}^2\left(\alpha+\alpha^{*}\right).
		\end{aligned}
		\label{motion_cavity}
	\end{equation}
	
	To validate the mean-field steady-state solutions, we numerically solve Eq.~\eqref{motion_cavity} across different phases by varying the parameters $\tilde{g}, \tau$. The resulting spin dynamics are visualized as trajectories projected onto the Bloch sphere, as shown in Fig.~\ref{fig:spin_dynamics}. For the solution branch with $s_z < 0$, the spin projections corresponding to the NP, SRP, and bistable phase are displayed  in Fig.~\ref{fig:spin_dynamics}(a--d).
    
    In the NP, the spin relaxes toward the normal steady state characterized by $s_z = -1$, as illustrated in Fig.~\ref{fig:spin_dynamics}(a). Within the SRP, the system evolves into a superradiant steady state with $-1 < s_z < 0$, shown in Fig.~\ref{fig:spin_dynamics}(b). In the bistable phase, the long‑time behavior depends sensitively on the initial conditions: the system may relax either to the normal steady state (Fig.~\ref{fig:spin_dynamics}(c)) or to the superradiant steady state (Fig.~\ref{fig:spin_dynamics}(d)).
    
    In addition, when the solution branch with $s_z > 0$ is selected, the system always evolves toward the spin‑inverted state characterized by $s_z = 1$, regardless of the initial conditions and the phase under consideration. This behavior is illustrated in Fig.~\ref{fig:spin_dynamics}(e--h).

    \subsection{Wigner function}
    
    \begin{figure}
		\centering		\includegraphics[width=0.45\textwidth]{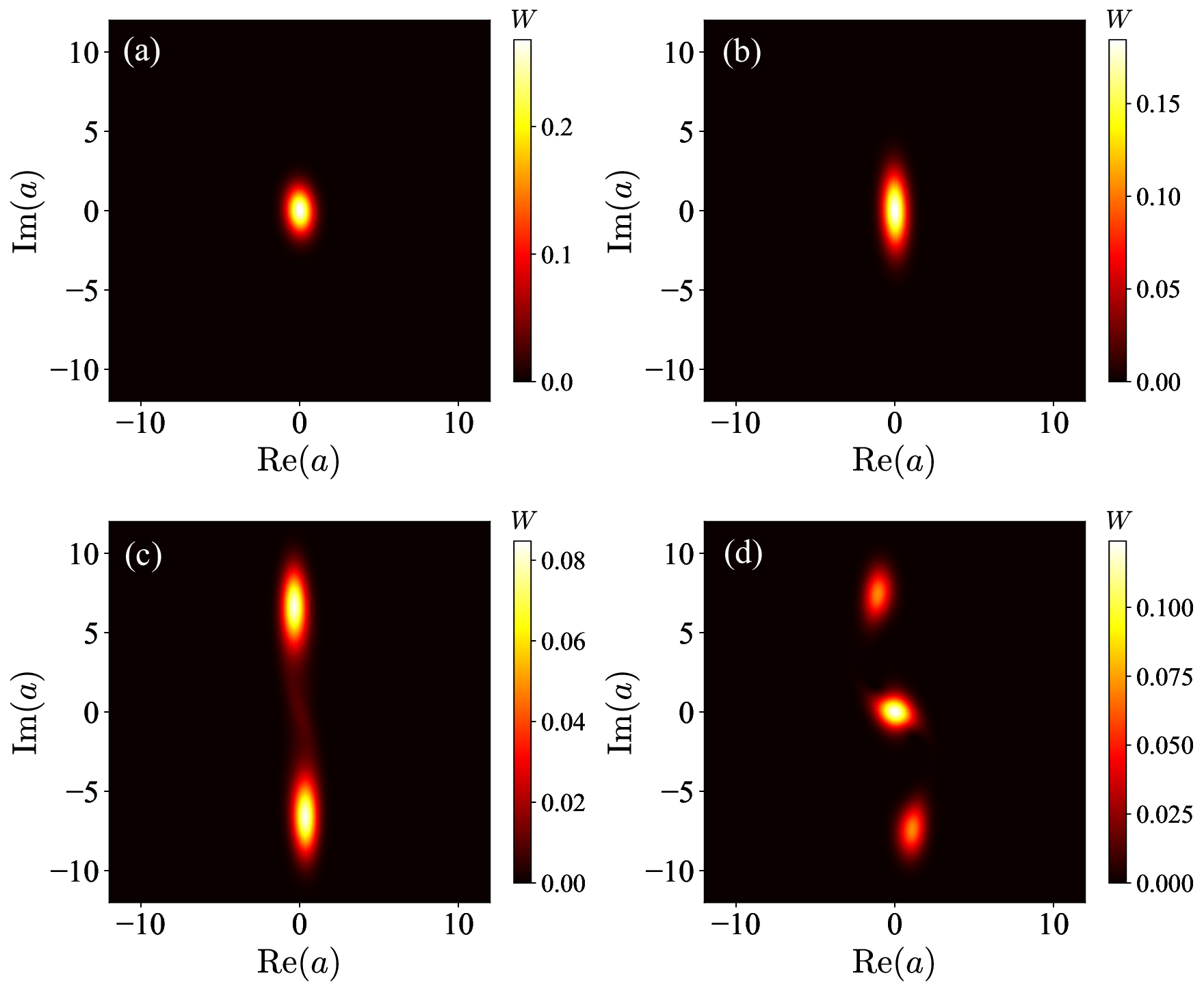}
		\caption{Steady-state Wigner functions of the cavity field for the anisotropic QRM including  the $A^2$ term. Each panel corresponds to a distinct phase: (a)–(b) NP,  (c) SRP, and (d) bistable phase. Upper panels (a)–(b) depict the symmetric case ($\tau = 1$) with $\tilde{g} = 0.5$ and $1$, respectively, where the superradiant phase transition is forbidden. Panels (c) and (d) correspond to the asymmetric case with $\tau = 2.4, \tilde{g} = 0.9$ and $\tau = 3.3, \tilde{g} = 0.6$, respectively. The Wigner functions are calculated with $\eta=\Delta/\omega=100$, Fock-space dimension $N=100$, $\kappa=3$, and $\gamma=0.5\omega$.}
		\label{fig:Wigner}
	\end{figure}
    
	The Wigner function provides a quasi‑probability distribution in phase space and offers a clear visualization of the field amplitude and its quantum coherence. In particular, Wigner function can indicate non‑classical features such as squeezing and interference. We display the Wigner function of the cavity-field in different steady‑state phases of the open anisotropic QRM, obtained by numerically projecting the steady state onto the spin‑down subspace at a finite frequency ratio. The single‑mode Wigner function is defined as
	\begin{equation}
		W(x_a,p_a) = \frac{1}{\pi}\int_{-\infty}^{\infty}dy^{\prime} e^{2ip_ay^{\prime}}\langle x_a-y^{\prime} \vert \rho \vert x_a+y^{\prime} \rangle,
	\end{equation}
	where $\rho$ is the reduced cavity density matrix obtained by projecting the full steady state onto the spin-down subspace $\ket{\downarrow}$.
	
    In the isotropic case $\tau=1$, the system remains in the NP for all coupling strengths. Consequently, the Wigner function displays a single elliptic peak centered at the origin, which is characteristic of a squeezed vacuum state, as shown in Fig.~\ref{fig:Wigner}(a--b). This behavior originates from the $A^2$ term, which opens an additional channel that suppresses  the onset of the superradiant phase transition. In the SRP, the two symmetry-related superradiant steady-state solutions give rise to a superposition of two displaced squeezed states, resulting in a squeezed Schr\"odinger-cat--like Wigner distribution, as illustrated in Fig.~\ref{fig:Wigner}(c). By contrast, in the bistable phase, the coexistence of NP and SRP solutions leads to a trimodal structure in the Wigner function, consisting of a central squeezed component associated with the NP and two symmetrically displaced squeezed components corresponding to the SRP, as shown in Fig.~\ref{fig:Wigner}(d). These results are converged with respect to the Fock-space cutoff for $\Delta/\omega=100$, and increasing the cutoff beyond $N=100$ does not lead to visible changes in the results.

	\section{\label{conclusion}conclusion}

     We have investigated the steady-state phase transitions of the open anisotropic QRM with an explicit $A^2$ term. The main effect of the $A^2$ term is to reshape the known dissipative phase diagram of the open anisotropic QRM. In particular, it breaks the symmetry of the phase structure with respect to  anisotropy, modifies the bistable regions, and changes the relation between the tricritical points and the merger of the two critical branches $\tilde g_c^\pm$.
     
     Through a comprehensive mean-field semiclassical analysis, we identify the emergence of three distinct phases: the NP, the SRP, and extended bistable phases characterized by the coexistence of NP and SRP. These three phases converge at tricritical points in the parameter space. With the exception of the first-order phase transition between the NP and the bistable phase, all other phase transitions are found to be of second order. The interplay between the anisotropy $\tau$ and the $A^2$ term strength $\kappa$ gives rise to modified DPT boundaries and enriches the resulting phase structure.
     
	The mean-field solutions are examined by linear stability analysis, while the corresponding cavity-field structures are characterized through Wigner function calculations. These results further illustrate the corresponding cavity-field structures and reveal squeezing features in the representative steady states.

   	The quantum fluctuations of the cavity field are further analyzed based on the effective Hamiltonians. Analytically, we find that near the intersection of the two critical couplings $\tilde{g}_c^{\pm}$, the photon-number fluctuations exhibit an anomalous critical exponent $\beta=-2$. This behavior differs from that at the tricritical point and contrasts with the case in which the  $A^2$ term is absent, demonstrating the high sensitivity of scaling properties to the presence of the $A^2$ term.  In earlier observations without the $A^2$ term, the anomalous exponent $\beta=-2$ near the tricritical point \cite{lyu2024PERL} could be attributed to the coincidence of the two critical couplings $\tilde{g}_c^{\pm}$  at that point. When $A^2$ term is introduced, the photon-number fluctuations near the tricritical point revert to the conventional scaling behavior with $\beta=-1$.

	This work provides an effective dissipative framework for analyzing the structure and properties of the steady-state phase diagram of the anisotropic QRM with an explicit $A^2$ term. It suggests possible directions for exploring dissipative multistability in engineered quantum platforms such as circuit QED and trapped-ion systems. The emergence of rich steady-state structures indicates that dissipation and the $A^2$ term can play important roles in shaping steady-state criticality, thereby providing a useful theoretical description of nonequilibrium phase transitions in related open quantum systems.
	
    \begin{acknowledgments}
     This work is supported by the National Key R\&D Program of China under Grant No. 2024YFA1408900 and the National Natural Science Foundation of China under Grant No. 92565201.
     \end{acknowledgments}

	\onecolumngrid
	\appendix
	\section{\label{appendix A}THE EXPLICIT SOLUTIONS FOR STEADY STATE}
    
Here, the nontrivial solutions of the steady-state equation Eq.~\eqref{steady} in maintext are presented, which can be rewritten as a system of four linear equations, $\mathcal{M}\left(x,y,s_x,s_y\right)^T = 0$, with
	\begin{equation}
		\mathcal{M} = 
		\begin{pmatrix}
			1+4\kappa \tilde{g}^2 &\tilde{\gamma} &\frac{\tilde{g}}{2}\left(1+\tau\right) &0\\
			\tilde{\gamma} &-1 &0 &\frac{\tilde{g}}{2}\left(1-\tau\right)\\
			-\tilde{g}\left(1+\tau\right)s_z &0 &\frac{1}{2} &0\\
			0 &\tilde{g}\left(1-\tau\right)s_z &0 &\frac{1}{2}
		\end{pmatrix}.
	\end{equation}
	For a linear equation system, the condition for nontrivial solutions is that the determinant vanishes, which leads to
	\begin{equation}
		\begin{aligned}
			\mathrm{det}\left[\mathcal{M}\right] &= -\frac{1}{4}\bigg\{2\left[\tilde{g}^2\left(1+\tau^2\right) +2\kappa\tilde{g}^4\left(1-\tau\right)^2 \right]s_z+\left(1+4\kappa\tilde{g}^2 + \tilde{\gamma}^2\right)+ \left(1-\tau^2\right)^2\tilde{g}^4s_z^2\bigg\}.
		\end{aligned}
	\end{equation}
	Then, the nontrivial solutions of $s_z$ are obtained as $s_z = -\left[ h\pm \sqrt{ h^2-q}\right]\big/\left[\left(1-\tau^2\right)^2\tilde{g}^2\right]$, where $h=\left[1 + \tau^2 + 2\kappa\tilde{g}^2(1-\tau)^2\right]$ and $q=(1-\tau^2)^2(1+4\kappa \tilde{g}^2+\tilde{\gamma}^2)$. Using the condition of spin conservation $s_x^2 + s_y^2 + s_z^2 = 1$ and the relation $s_x = 2\tilde{g}\left(1+\tau\right)xs_z, s_y = 2\tilde{g}\left(\tau - 1\right)ys_z$, the corresponding nontrivial solutions for $x, y$ are
	\begin{equation}
		\begin{aligned}
			x =  \pm \sqrt{\frac{1-s_z^2}{4\tilde{g}^2s_z^2\left[\left(1+\tau\right)^2 +\frac{\tilde{\gamma}^2\left(1-\tau\right)^2}{\left[1+\tilde{g}^2\left(1-\tau\right)^2s_z\right]^2}\right]}},\quad	y =  \pm \sqrt{\frac{1-s_z^2}{4\tilde{g}^2s_z^2\left[\left(1-\tau\right)^2 +\frac{\tilde{\gamma}^2\left(1+\tau\right)^2}{\left[1+4\kappa\tilde{g}^2+\tilde{g}^2\left(1+\tau\right)^2s_z\right]^2}\right]}}.
		\end{aligned}
	\end{equation}

   \begin{figure*}
    \centering
    \includegraphics[width=0.9\textwidth]{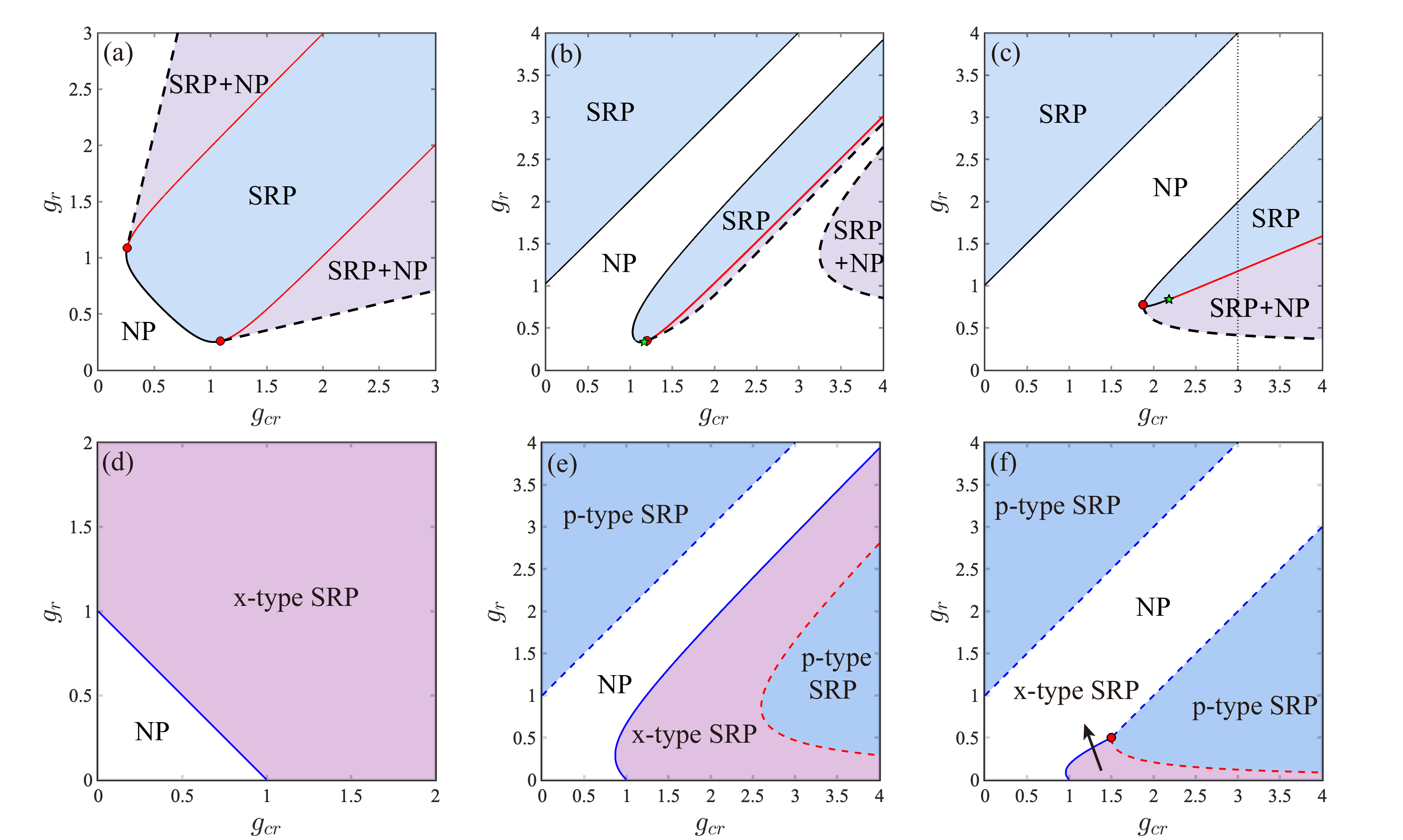}
    \caption{The phase diagram from Fig. \ref{fig:phase_diagram} is replotted in the $(g_{cr}=\tau\tilde{g}>0,\ g_{r}=\tilde{g}>0)$-plane, keeping the same notation as in the original figure. In panel (c), the black dotted line traces a path at fixed CRW coupling strength $g_{cr}=3$, which is used in the main text to analyze fluctuations and universal critical behavior.}
    \label{fig:phase_diagram1}
    \end{figure*}
    
   In this appendix, we replot the phase diagram of both the open and closed QRMs (Fig.~\ref{fig:phase_diagram}) in the $(g_{cr}=\tau\tilde{g}>0,\ g_{r}=\tilde{g}>0)$-plane as Fig.~\ref{fig:phase_diagram1} for reference in the main text. In this new representation, $\tau$ is always positive. Notably, in the absence of the $A^2$ term ($\kappa = 0$), the phase diagrams for both the open and closed QRMs exhibit symmetry about the diagonal line $\tau = 1$. With the inclusion of the $A^2$ term, however, they become asymmetric.

	\section{\label{appendix B}ADIABATICALLY ELIMINATION IN STABILITY ANALYSIS}
	Here, details of the stability analysis for the mean-field solutions in the frequency limit $\Delta/\omega \rightarrow \infty$ are presented. After expanding the order parameters around their mean-field values, $\alpha \rightarrow \alpha + \delta\alpha$, $s_- \rightarrow s_- + \delta s_-$, and $s_z \rightarrow s_z + \delta s_z$, the equations of motion for these fluctuations can be expressed as
	\begin{equation}
		\begin{aligned}
			\frac{d}{dt}\delta x &= - \omega \tilde{\gamma} \delta x + \omega \delta y - \omega\frac{\left(1-\tau\right)\tilde{g}}{2} \delta s_y,\\
			\frac{d}{dt}\delta y &=-\omega\left(1+4\kappa\tilde{g}^2\right)\delta x - \omega\tilde{\gamma}\delta y - \omega\frac{\left(1+\tau\right)\tilde{g}}{2}\delta s_x,\\
			\frac{d}{dt}\delta s_x &= -2\Delta \left(1-\tau\right)\tilde{g}s_z \delta y + 2\Delta \left(1-\tau\right)\tilde{g} y \frac{s_x}{s_z}\delta s_x+ \Delta\left[2\left(1-\tau\right)\tilde{g}y\frac{s_y}{x_z} - 1\right]\delta s_y,\\
			\frac{d}{dt}\delta s_y &= -2\Delta\left(1+\tau\right)\tilde{g}s_z \delta x +\Delta\left[2\left(1+\tau\right)\tilde{g}x\frac{s_x}{x_z}+1\right]\delta s_x + 2\Delta\left(1+\tau\right)\tilde{g}x\frac{s_y}{s_z} \delta s_y.
		\end{aligned}
		\label{stability}
	\end{equation}
	Then, the relation between the spin fluctuations is used as $\delta s_z = -\frac{1}{s_z}\left(s_x\,\delta s_x + s_y\,\delta s_y\right) = -2\tilde{g}(1+\tau)x\,\delta s_x + 2\tilde{g}(1-\tau)y\,\delta s_y$. By rewriting Eq.~\eqref{stability} in matrix form, the dynamical equations of the fluctuations become
		\begin{equation}
			\frac{d}{dt}\begin{pmatrix}
				\delta x \\ \delta y \\ \delta s_x \\ \delta s_y
			\end{pmatrix} = \begin{pmatrix}
				-\omega \tilde{\gamma} &\omega &0 &-\omega\frac{\left(1-\tau\right)\tilde{g}}{2}\\
				-\omega\left(1+4\kappa\tilde{g}^2\right) &-\omega \tilde{\gamma} &-\omega \frac{\left(1+\tau\right)\tilde{g}}{2} &0\\
				0 & -2\Delta\left(1-\tau\right)\tilde{g}s_z &2\Delta \left(1-\tau\right)\tilde{g} y \frac{s_x}{s_z} &\Delta\left[2\left(1-\tau\right)\tilde{g}y\frac{s_y}{x_z} - 1\right]\\
				-2\Delta\left(1+\tau\right)\tilde{g}s_z &0 &\Delta\left[2\left(1+\tau\right)\tilde{g}x\frac{s_x}{x_z}+1\right] &2\Delta\left(1+\tau\right)\tilde{g}x\frac{s_y}{s_z}
			\end{pmatrix}\begin{pmatrix}
				\delta x \\
				\delta y\\
				\delta s_x\\
				\delta s_y
			\end{pmatrix}.
			\label{stability_matrix}
		\end{equation}
		The spin fluctuations can be eliminated in the frequency limit $\Delta/\omega \rightarrow \infty$ by imposing $d\delta s_x/dt = 0$ and $d\delta s_y/dt = 0$.  Substituting Eq.~\eqref{adiabation} into Eq.~\eqref{stability_matrix} , we obtain the reduced dynamical equation for the field fluctuations only
		\begin{equation}
			\frac{d}{dt}\begin{pmatrix}
				\delta x \\ \delta y 
			\end{pmatrix} = \omega \mathbb{M}\begin{pmatrix}
				\delta x\\
				\delta y
			\end{pmatrix},
			\label{stability_matrix_re}
		\end{equation}
		where the elements of the stability matrix $\mathbb{M}$ are
		\begin{equation}
			\begin{aligned}
				\mathbb{M}_{11} &= -\tilde{\gamma} - {4s_z x y \left(1-\tau^2\right)^2\tilde{g}^4}/{\Sigma},\quad
				\mathbb{M}_{12} = 1 + {s_z\left(1-\tau\right)^2\left[4x^2\left(1+\tau\right)^2\tilde{g}^2 + 1\right]\tilde{g}^2}/{\Sigma},\\
				\mathbb{M}_{21} &= -\left(1 + 4\kappa\tilde{g}^2\right) - {s_z\left(1+\tau\right)^2\left[4y^2\left(1-\tau\right)^2\tilde{g}^2 + 1\right]\tilde{g}^2}/{\Sigma},\quad \mathbb{M}_{22} =  -\tilde{\gamma} + {4s_z x y \left(1-\tau^2\right)^2\tilde{g}^4}/{\Sigma}.
			\end{aligned}
		\end{equation}
		
	\section{\label{appendix C}THE EFFECTIVE HAMILTONIAN FOR SUPERRADIANT PHASE}
	The detailed derivation for obtaining the effective Hamiltonian
	of the SRP is presented. First, a displacement unitary transformation
	is applied to the original master equation \eqref{lindblad} with the
	displacement operator $D\left(\alpha\right) = \exp\left[\sqrt{\Delta/\omega}\left(\alpha a^\dagger - \alpha^* a\right)\right]$. The transformed master equation is then given by
	\begin{equation}
		\dot{\bar{\rho}} =  -i\left[\bar{H},\bar{\rho}\right] + \gamma \mathcal{D}\left[{a}\right],
	\end{equation}
	where $\bar{\rho} = D^\dagger \rho D$, $\bar{H} = D^\dagger H D + i\sqrt{\Delta \omega}\tilde{\gamma}\left(\alpha^* a -\alpha a^\dagger\right)
	$. The transformed Hamiltonian reads
	\begin{equation}
		\begin{aligned}
			\bar{H} &= \omega a^\dagger a + \omega \kappa \tilde{g}^2 \left(a + a^\dagger\right)^2 + \Delta \left(|\alpha|^2 + 4\kappa\tilde{g}^2 x^2\right)+\frac{\Delta}{2}\sigma_z + \Delta \tilde{g}\left[\left(\alpha^* + \tau \alpha\right)\sigma_{+} + h.c. \right]\\
			& + \sqrt{\Delta \omega} \tilde{g}\left[\left(a\sigma_{+} + a^\dagger \sigma_{-}\right) + h.c. \right]+ \sqrt{\Delta \omega}\left\{\left[\left(1+i\tilde{\gamma}\right)\alpha^* + 4\kappa\tilde{g}^2x\right]a + h.c.\right\}. 
		\end{aligned}
		\label{Hamiltonian_dis}
	\end{equation} 
	The term of the Hamiltonian \eqref{Hamiltonian_dis}, $\bar{H}_s$, describes only the spin part and can be diagonalized as $U^\dagger \bar{H}_s U = \frac{\Delta}{2} \sqrt{1+4\tilde{g}^2|\alpha + \tau \alpha^*|^2}\,\sigma_z = \frac{\Delta\sigma_z}{2|s_z|}$ by the unitary operator
	\begin{equation}
		U = \frac{1}{\sqrt{2}\sqrt{\chi+\sqrt{\chi}}}\begin{pmatrix}
			1 + \sqrt{\chi} &-2\tilde{g}\left(\alpha^* + \tau \alpha\right)\\
			2\tilde{g}\left(\alpha + \tau \alpha^*\right) &1+\sqrt{\chi}
		\end{pmatrix},
	\end{equation}
	where $\chi = 1/s_z^2$. Under the unitary transformation defined above, the interaction term between the atom and the field, $\bar{H}_{\mathrm{int}}$, takes the form
	\begin{equation}
		\begin{aligned}
			& U^\dagger \bar{H}_{\text{int}} U =  \sqrt{\Delta \omega} \left[\left(\bar{g}_r a\sigma_{+} + \bar{g}^*_r a^\dagger \sigma_{-}\right) + \left(\bar{g}_{cr}a^\dagger \sigma_{+} + \bar{g}_{cr}^* a \sigma_{-}\right)\right]-\sqrt{\Delta \omega}\left(\eta a^\dagger + \eta^*a\right)\sigma_z,
		\end{aligned}
	\end{equation}
	with 
	\begin{equation}
		\begin{aligned}
			\bar{g}_r &=-\frac{\tilde{g}}{2} \left[1+\frac{1}{\sqrt{\chi}}-4 \tau \tilde{g}^2\frac{\left(\alpha + \tau \alpha^*\right)^2}{\chi + \sqrt{\chi}}\right],\\
			\bar{g}_{cr} &= -\frac{\tilde{g}}{2}\left[\tau\left(1+\frac{1}{\sqrt{\chi}}\right) -4 \tilde{g}^2\frac{\left(\alpha + \tau \alpha^*\right)^2}{\chi + \sqrt{\chi}}\right],\\
			\eta = & \frac{\tilde{g}^2 }{\sqrt{\chi}}\left[\left(1+\tau^2\right)\alpha + 2\tau \alpha^*\right].
		\end{aligned}
	\end{equation}
	Thus, the Hamiltonian \eqref{Hamiltonian_dis} in the new spin basis, $\bar{H}^{\prime}=U^\dagger \bar{H} U$, can be rewritten as
	\begin{equation}
		\begin{aligned}
			\bar{H}^\prime &= 	\omega a^\dagger a + \omega \kappa \tilde{g}^2 \left(a + a^\dagger\right)^2 + \Delta \left(|\alpha|^2 + 4\kappa\tilde{g}^2 x^2\right) + \frac{\Delta\sigma_z}{2|s_z|}\\
			& + \sqrt{\Delta 	\omega} \left[\left(\bar{g}_r a\sigma_{+} + \bar{g}^*_r a^\dagger \sigma_{-}\right) + \left(\bar{g}_{cr}a^\dagger \sigma_{+} + \bar{g}_{cr}^* a \sigma_{-}\right)\right],
		\end{aligned}
	\end{equation}
	It can be readily verified that the linear terms in $\bar{H}^{\prime}$
	vanish after projection onto the spin-down subspace $\ket{\downarrow}$. Subsequently, we preform a squeezing transformation $\bar{H}^{\prime \prime}=S^\dagger \bar{H}^\prime S$ to eliminate $A^2$ term by squeezing operator $S=\exp\left[\frac{r}{2}\left(a^{\dagger 2}-a^2\right)\right]$ with $r=-\frac{1}{4}\ln\left(1+4\kappa\tilde{g}^2\right)$. The transformed Hamiltonian is
	\begin{equation}
		\begin{aligned}
			\bar{H}^{\prime \prime} &=  \omega^{\prime} a^\dagger a + \frac{\Delta^{\prime}}{2}\sigma_z+ \sqrt{\Delta \omega} \left(\bar{g}_r^\prime a \sigma_+ +  \bar{g}_r^{\prime *} a^\dagger \sigma_{-} + \bar{g}_{cr}^\prime a^\dagger \sigma_+ + \bar{g}_{cr}^{\prime *} a\sigma_{-}\right),
		\end{aligned}
	\end{equation}
	where $\bar{g}_r^\prime = \bar{g}_r\cosh r + \bar{g}_{cr}\sinh r$, $\bar{g}_{cr}^\prime = \bar{g}_r \sinh r + \bar{g}_{cr}\cosh r$ and $\Delta^{\prime}=\Delta/|s_z|$. A SW transformation  $H_{\text{SRP}} = V^\dagger \bar{H}^{\prime \prime} V$ is then applied, with $V = \frac{\sqrt{\Delta \omega}}{\Delta^\prime} \left(\bar{g}_r^{\prime*}a^\dagger \sigma_{-} - \bar{g}_r^{\prime}a\sigma_{+} + \bar{g}_{cr}^{\prime*}a \sigma_{-}- \bar{g}_{cr}^{\prime}a^\dagger \sigma_{+} \right)$. Finally, projecting $H_{\text{SRP}}$ onto the spin-down subspace
	$\ket{\downarrow}$ yields the effective Hamiltonian for the SRP,
	\begin{equation}
		\begin{aligned}
			H_{\text{SRP}}^{\downarrow} &= R_{\text{SRP}}a^\dagger a + P_{\text{SRP}}a^{\dagger 2} + P_{\text{SRP}}^* a^{2},
		\end{aligned}
	\end{equation}
	where $R_{\text{SRP}} = \omega\left[\xi - |s_z|\left(|\bar{g}_r^\prime|^2 + |\bar{g}_{cr}^\prime|^2\right)\right]$ and $P_{\text{SRP}} = -\omega |s_z|\bar{g}_r^{\prime} \bar{g}_{cr}^{\prime*}$. The effective master equation for SRP become $\dot{\rho}_{\text{SRP}}=-i\left[H_{\text{SRP}}^{\downarrow},\rho_{\text{SRP}}\right] + \gamma \mathcal{D}\left[\bar{a}\right]$, where $\rho_{\text{SRP}}= \bra{\downarrow} V^\dagger S^\dagger U^\dagger \bar{\rho} U S V\ket{\downarrow}$ and $\bar{a} = V^\dagger S^\dagger U^\dagger a U S V = a\cosh r + a^\dagger \sinh r$.

    \twocolumngrid
    \bibliography{main}

@article{DM,
		author = {R. H. {Dicke}},
		title  = {Coherence in spontaneous radiation processes},
		journal= {Phys. Rev.},
		volume = {93},
		pages  = {99},
		year   = {1954}
	}

@article{DM_PT,
		author = {K. Hepp and E. H. Lieb},
		title  = {On the superradiant phase transition for molecules in a quantized radiation field: the {Dicke} maser model},
		journal= {Ann. Phys. (N.Y.)},
		volume = {76},
		pages  = {360},
		year   = {1973}
	}

@article{DM_PT2,
		author = {Y. K. Wang and F. T. Hioe},
		title  = {Phase transition in the {Dicke} model of superradiance},
		journal= {Phys. Rev. A},
		volume = {7},
		pages  = {831},
		year   = {1973}
	}

@article{emary2003prl,
		author = {C. Emary and T. Brandes},
		title  = {Quantum chaos triggered by precursors of a quantum phase transition: The {Dicke} model},
		journal= {Phys. Rev. Lett.},
		volume = {90},
		pages  = {044101},
		year   = {2003}
	}

@article{tniemczyk2010np,
		author = {T. Niemczyk and F. Deppe and H. Huebl and E. P. Menzel and F. Hocke and M. J. Schwarz and J. J. Garcia-Ripoll and D. Zueco and T. H\"ummer and E. Solano and A. Marx and R. Gross},
		title  = {Circuit quantum electrodynamics in the ultrastrong-coupling regime},
		journal= {Nat. Phys.},
		volume = {6},
		pages  = {772},
		year   = {2010}
	}

@article{forn2010prl,
		author = {P. Forn-D\'{\i}az and J. Lisenfeld and D. Marcos and J. J. Garc\'{\i}a-Ripoll and E. Solano and C. J. P. M. Harmans and J. E. Mooij},
		title  = {Observation of the {Bloch--Siegert} shift in a qubit-oscillator system in the ultrastrong coupling regime},
		journal= {Phys. Rev. Lett.},
		volume = {105},
		pages  = {237001},
		year   = {2010}
	}

@article{yoshihara2016np,
		author = {F. Yoshihara and T. Fuse and S. Ashhab and K. Kakuyanagi and S. Saito and K. Semba},
		title  = {Superconducting qubit-oscillator circuit beyond the ultrastrong-coupling regime},
		journal= {Nat. Phys.},
		volume = {13},
		pages  = {44},
		year   = {2017}
	}

@article{trap_ions,
		author = {J. S. Pedernales and I. Lizuain and S. Felicetti and G. Romero and L. Lamata and E. Solano},
		title  = {Quantum {Rabi} Model with Trapped Ions},
		journal= {Sci. Rep.},
		volume = {5},
		pages  = {15472},
		year   = {2015}
	}

@article{lmduan2021nc,
		author = {M.-L. Cai and Z.-D. Liu and W.-D. Zhao and Y.-K. Wu and Q.-X. Mei and Y. Jiang and L. He and X. Zhang and Z.-C. Zhou and L.-M. Duan},
		title  = {Observation of a quantum phase transition in the quantum {Rabi} model with a single trapped ion},
		journal= {Nat. Commun.},
		volume = {12},
		pages  = {1126},
		year   = {2021}
	}

@article{cold_atom,
		author = {S. Felicetti and E. Rico and C. Sabin and T. Ockenfels and J. Koch and M. Leder and C. Grossert and M. Weitz and E. Solano},
		title  = {Quantum {Rabi} model in the Brillouin zone with ultracold atoms},
		journal= {Phys. Rev. A},
		volume = {95},
		pages  = {013827},
		year   = {2017}
	}

@article{cuiti2014prl,
		author = {A. Baksic and C. Ciuti},
		title  = {Controlling discrete and continuous symmetries in ``superradiant'' phase transitions with circuit {QED} systems},
		journal= {Phys. Rev. Lett.},
		volume = {112},
		pages  = {173601},
		year   = {2014}
	}

@article{keeling2019aqt,
		author = {P. Kirton and M. M. Roses and J. Keeling and E. G. Dalla Torre},
		title  = {Introduction to the {{Dicke}} model: From equilibrium to nonequilibrium, and vice versa},
		journal= {Adv. Quantum Technol.},
		volume = {2},
		pages  = {1800043},
		year   = {2019}
	}

@article{hwang2015prl,
		author = {M.-J. Hwang and R. Puebla and M. B. Plenio},
		title  = {Quantum phase transition and universal dynamics in the {Rabi} model},
		journal= {Phys. Rev. Lett.},
		volume = {115},
		pages  = {180404},
		year   = {2015}
	}

@article{QRM,
		author = {I. I. Rabi},
		title  = {Space quantization in a gyrating magnetic field},
		journal= {Phys. Rev.},
		volume = {51},
		pages  = {652},
		year   = {1937}
	}

@article{hwang2016prl,
		author = {M.-J. Hwang and M. B. Plenio},
		title  = {Quantum phase transition in the finite {J}aynes--{C}ummings lattice systems},
		journal= {Phys. Rev. Lett.},
		volume = {117},
		pages  = {123602},
		year   = {2016}
	}

@article{mxliu2017prl,
		author = {M. Liu and S. Chesi and Z.-J. Ying and X. Chen and H.-G. Luo and H.-Q. Lin},
		title  = {Universal scaling and critical exponents of the anisotropic quantum {Rabi} model},
		journal= {Phys. Rev. Lett.},
		volume = {119},
		pages  = {220601},
		year   = {2017}
	}

@article{shen2017pra,
		author = {L.-T. Shen and Z.-B. Yang and H.-Z. Wu and S.-B. Zheng},
		title  = {Quantum phase transition and quench dynamics in the anisotropic {Rabi} model},
		journal= {Phys. Rev. A},
		volume = {95},
		pages  = {013819},
		year   = {2017}
	}

@article{zhang2021prl,
		author = {Y.-Y. Zhang and Z.-X. Hu and L. Fu and H.-G. Luo and H. Pu and X.-F. Zhang},
		title  = {Quantum phases in a quantum {Rabi} triangle},
		journal= {Phys. Rev. Lett.},
		volume = {127},
		pages  = {063602},
		year   = {2021}
	}

@article{A2original,
		author = {K. Rza\.zewski and K. W\'odkiewicz and W. \.{Z}akowicz},
		title  = {Phase transitions, two-level atoms, and the $\mathbf{A}^{2}$ term},
		journal= {Phys. Rev. Lett.},
		volume = {35},
		pages  = {432},
		year   = {1975}
	}

@article{keeling2007jpb,
		author = {J. Keeling},
		title  = {Coulomb interactions, gauge invariance, and phase transitions of the {Dicke} model},
		journal= {J. Phys. Condens. Matter},
		volume = {19},
		pages  = {295213},
		year   = {2007}
	}

@article{vukics2014prl,
		author = {A. Vukics and T. Grie{\ss}er and P. Domokos},
		title  = {Elimination of the {A-square} problem from cavity {QED}},
		journal= {Phys. Rev. Lett.},
		volume = {112},
		pages  = {073601},
		year   = {2014}
	}

@article{ciuti2010nc,
		author = {P. Nataf and C. Ciuti},
		title  = {No-go theorem for superradiant quantum phase transitions in cavity {QED} and counter-example in circuit {QED}},
		journal= {Nat. Commun.},
		volume = {1},
		pages  = {72},
		year   = {2010}
	}

@article{viehmann2011prl,
		author = {O. Viehmann and J. von Delft and F. Marquardt},
		title  = {Superradiant phase transitions and the standard description of circuit {QED}},
		journal= {Phys. Rev. Lett.},
		volume = {107},
		pages  = {113602},
		year   = {2011}
	}

@article{andolina2019prb,
		author = {G. M. Andolina and F. M. D. Pellegrino and V. Giovannetti and A. H. MacDonald and M. Polini},
		title  = {Cavity quantum electrodynamics of strongly correlated electron systems: A no-go theorem for photon condensation},
		journal= {Phys. Rev. B},
		volume = {100},
		pages  = {121109(R)},
		year   = {2019}
	}

@article{nazir2020prl,
		author = {A. Stokes and A. Nazir},
		title  = {Uniqueness of the phase transition in many-dipole cavity quantum electrodynamical systems},
		journal= {Phys. Rev. Lett.},
		volume = {125},
		pages  = {143603},
		year   = {2020}
	}

@article{ciuti2013pra,
		author = {A. Baksic and P. Nataf and C. Ciuti},
		title  = {Superradiant phase transitions with three-level systems},
		journal= {Phys. Rev. A},
		volume = {87},
		pages  = {023813},
		year   = {2013}
	}

@article{taoliu2012pla,
		author = {T. Liu and Y. Y. Zhang and Q. H. Chen and K. L. Wang},
		title  = {Possibility of superradiant phase transitions in coupled two-level atoms},
		journal= {Phys. Lett. A},
		volume = {376},
		pages  = {1962},
		year   = {2012}
	}

@article{ymwang2020pra,
		author = {Y. Wang and M. Liu and W.-L. You and S. Chesi and H.-G. Luo and H.-Q. Lin},
		title  = {Resilience of the superradiant phase against $A^{2}$ effects in the quantum {Rabi} dimer},
		journal= {Phys. Rev. A},
		volume = {101},
		pages  = {063843},
		year   = {2020}
	}

@article{Peng,
		author = {X. Chen and Z. Wu and M. Jiang and X.-Y. L\"u and X. Peng and J. Du},
		title  = {Experimental quantum simulation of superradiant phase transition beyond no-go theorem via antisqueezing},
		journal= {Nat. Commun.},
		volume = {12},
		pages  = {6281},
		year   = {2021}
	}

@article{xychen,
		author = {X.-Y. Chen and L. Duan and D. Braak and Q.-H. Chen},
		title  = {Multiple ground-state instabilities in the anisotropic quantum {Rabi} model},
		journal= {Phys. Rev. A},
		volume = {103},
		pages  = {043708},
		year   = {2021}
	}

@article{qtxie2014prx,
		author = {Q.-T. Xie and S. Cui and J.-P. Cao and L. Amico and H. Fan},
		title  = {Anisotropic {Rabi} model},
		journal= {Phys. Rev. X},
		volume = {4},
		pages  = {021046},
		year   = {2014}
	}

@article{erlingsson2010prb,
		author = {S. I. Erlingsson and J. C. Egues and D. Loss},
		title  = {Energy spectra for quantum wires and two-dimensional electron gases in magnetic fields with Rashba and Dresselhaus spin-orbit interactions},
		journal= {Phys. Rev. B},
		volume = {82},
		pages  = {155456},
		year   = {2010}
	}

@article{skogvoll2021pra,
		author = {I. C. Skogvoll and J. Lidal and J. Danon and A. Kamra},
		title  = {Tunable anisotropic quantum {Rabi} model via a magnon-spin-qubit Ensemble},
		journal= {Phys. Rev. Appl.},
		volume = {16},
		pages  = {064008},
		year   = {2021}
	}

@article{Dicke_Stark,
		author = {X.-Y. Chen and Y.-Y. Zhang and Q.-H. Chen and H.-Q. Lin},
		title  = {Phase transitions in the anisotropic {Dicke--Stark} model with $A^{2}$ terms},
		journal= {Phys. Rev. A},
		volume = {110},
		pages  = {063722},
		year   = {2024}
	}

@article{ye2025PRA,
		author = {T. Ye and Y.-Z. Wang and X.-Y. Chen and Q.-H. Chen and H.-Q. Lin},
		title  = {Superradiant phase transitions in the quantum {Rabi} model: Overcoming the no-go theorem through anisotropy},
		journal= {Phys. Rev. A},
		volume = {111},
		pages  = {043716},
		year   = {2025}
	}

@article{Hwang2018PRA,
		author = {M.-J. Hwang and P. Rabl and M. B. Plenio},
		title  = {Dissipative phase transition in the open quantum {Rabi} model},
		journal= {Phys. Rev. A},
		volume = {97},
		pages  = {013825},
		year   = {2018}
	}

@article{lyu2024PERL,
		author = {G. T. Lyu and K. Kottmann and M. B. Plenio and M.-J. Hwang},
		title  = {Multicritical dissipative phase transitions in the anisotropic open quantum {Rabi} model},
		journal= {Phys. Rev. Res.},
		volume = {6},
		pages  = {033075},
		year   = {2024}
	}

@article{rossini2021pr,
		author = {D. Rossini and E. Vicari},
		title  = {Coherent and dissipative dynamics at quantum phase transitions},
		journal= {Phys. Rep.},
		volume = {936},
		pages  = {1--110},
		year   = {2021}
	}

@article{bartolo2016pra,
		author = {N. Bartolo and F. Minganti and W. Casteels and C. Ciuti},
		title  = {Exact steady state of a Kerr resonator with one- and two-photon driving and dissipation: Controllable Wigner-function multimodality and dissipative phase transitions},
		journal= {Phys. Rev. A},
		volume = {94},
		pages  = {033841},
		year   = {2016}
	}

@article{krimer2019prl,
		author = {D. O. Krimer and M. Pletyukhov},
		title  = {Few-Mode Geometric Description of a Driven-Dissipative Phase Transition in an Open Quantum System},
		journal= {Phys. Rev. Lett.},
		volume = {123},
		pages  = {110604},
		year   = {2019}
	}

@article{rodriguez2017prl,
		author = {S. R. K. Rodriguez and W. Casteels and F. Storme and N. Carlon Zambon and I. Sagnes and L. Le Gratiet and E. Galopin and A. Lema\^{\i}tre and A. Amo and C. Ciuti and J. Bloch},
		title  = {Probing a dissipative phase transition via dynamical optical hysteresis},
		journal= {Phys. Rev. Lett.},
		volume = {118},
		pages  = {247402},
		year   = {2017}
	}

@article{tomita2017sciadv,
		author = {T. Tomita and S. Nakajima and I. Danshita and Y. Takasu and Y. Takahashi},
		title  = {Observation of the Mott insulator to superfluid crossover of a driven-dissipative Bose--Hubbard system},
		journal= {Sci. Adv.},
		volume = {3},
		pages  = {e1701513},
		year   = {2017}
	}

@article{fitzpatrick2017prx,
		author = {M. Fitzpatrick and N. M. Sundaresan and A. C. Y. Li and J. Koch and A. A. Houck},
		title  = {Observation of a dissipative phase transition in a one-dimensional circuit {QED} lattice},
		journal= {Phys. Rev. X},
		volume = {7},
		pages  = {011016},
		year   = {2017}
	}

@article{brennecke2013pnas,
		author = {F. Brennecke and R. Mottl and K. Baumann and R. Landig and T. Donner and T. Esslinger},
		title  = {Real-time observation of fluctuations at the driven-dissipative {Dicke} phase transition},
		journal= {Proc. Natl. Acad. Sci. U.S.A.},
		volume = {110},
		pages  = {11763},
		year   = {2013}
	}

@article{klinder2015pnas,
		author = {J. Klinder and H. Ke\ss ler and M. Wolke and L. Mathey and A. Hemmerich},
		title  = {Dynamical phase transition in the open {Dicke} model},
		journal= {Proc. Natl. Acad. Sci. U.S.A.},
		volume = {112},
		pages  = {3290},
		year   = {2015}
	}

@article{kollar2017nc,
		author = {A. J. Koll\'ar and A. T. Papageorge and V. D. Vaidya and Y. Guo and J. Keeling and B. L. Lev},
		title  = {Supermode--density--wave--polariton condensation with a Bose--Einstein condensate in a multimode cavity},
		journal= {Nat. Commun.},
		volume = {8},
		pages  = {14386},
		year   = {2017}
	}

@article{cai2022cpl,
		author = {M.-L. Cai and Z.-D. Liu and Y. Jiang and Y.-K. Wu and Q.-X. Mei and W.-D. Zhao and L. He and X. Zhang and Z.-C. Zhou and L.-M. Duan},
		title  = {Probing a dissipative phase transition with a trapped ion through reservoir engineering},
		journal= {Chin. Phys. Lett.},
		volume = {39},
		pages  = {020502},
		year   = {2022}
	}

@article{zhao2025prl,
		author = {X. Zhao and Q. Bin and W. Hou and Y. Li and Y. Li and Y. Lin and X.-Y. L\"u and J. Du},
		title  = {Experimental observation of parity-symmetry-protected phenomena in the quantum {Rabi} model with a trapped ion},
		journal= {Phys. Rev. Lett.},
		volume = {134},
		pages  = {193604},
		year   = {2025}
	}

@article{ashhab2013superradiance,
      title={Superradiance transition in a system with a single qubit and a single oscillator},
      author={Ashhab, S},
      journal={Phys. Rev. A},
      volume={87},
      number={1},
      pages={013826},
      year={2013},
      publisher={APS}
    }

@article{riedel1974effective,
      title={Effective critical and tricritical exponents},
      author={Riedel, Eberhard K and Wegner, Franz J},
      journal={Phys. Rev. B},
      volume={9},
      number={1},
      pages={294},
      year={1974},
      publisher={APS}
    }

@article{torre2013keldysh,
      title={Keldysh approach for nonequilibrium phase transitions in quantum optics: Beyond the {Dicke} model in optical cavities},
      author={Torre, Emanuele G Dalla and Diehl, Sebastian and Lukin, Mikhail D and Sachdev, Subir and Strack, Philipp},
      journal={Physical Review A—Atomic, Molecular, and Optical Physics},
      volume={87},
      number={2},
      pages={023831},
      year={2013},
      publisher={APS}
    }

@article{keeling2010collective,
      title={Collective dynamics of Bose-Einstein condensates in optical cavities},
      author={Keeling, J and Bhaseen, MJ and Simons, BD},
      journal={Physical review letters},
      volume={105},
      number={4},
      pages={043001},
      year={2010},
      publisher={APS}
    }

@article{rouse2023theory,
      title={Theory of photon condensation in an arbitrary-gauge condensed matter cavity model},
      author={Rouse, Dominic M and Stokes, Adam and Nazir, Ahsan},
      journal={Phys. Rev. B},
      volume={107},
      number={20},
      pages={205128},
      year={2023},
      publisher={APS}
    }

@article{minganti2018spectral,
    	title={Spectral theory of Liouvillians for dissipative phase transitions},
    	author={Minganti, Fabrizio and Biella, Alberto and Bartolo, Nicola and Ciuti, Cristiano},
    	journal={Phys. Rev. A},
    	volume={98},
    	number={4},
    	pages={042118},
    	year={2018},
    	publisher={APS}
    }

@article{le2013steady,
    	title={Steady-State Phases and Tunneling-Induced Instabilities in the Driven Dissipative Bose-Hubbard Model},
    	author={Le Boit{\'e}, Alexandre and Orso, Giuliano and Ciuti, Cristiano},
    	journal={Phys. Rev. Lett.},
    	volume={110},
    	number={23},
    	pages={233601},
    	year={2013},
    	publisher={APS}
    }

@article{le2014bose,
    	title={Bose-Hubbard model: Relation between driven-dissipative steady states and equilibrium quantum phases},
    	author={Le Boit{\'e}, Alexandre and Orso, Giuliano and Ciuti, Cristiano},
    	journal={Phys. Rev. A},
    	volume={90},
    	number={6},
    	pages={063821},
    	year={2014},
    	publisher={APS}
    }

@article{carr2013nonequilibrium,
    	title={Nonequilibrium phase transition in a dilute {Rydberg} ensemble},
    	author={Carr, Christopher and Ritter, Ralf and Wade, Christopher G and Adams, Charles S and Weatherill, Kevin J},
    	journal={Phys. Rev. Lett.},
    	volume={111},
    	number={11},
    	pages={113901},
    	year={2013},
    	publisher={APS}
    }

@article{marcuzzi2014universal,
    	title={Universal nonequilibrium properties of dissipative {Rydberg gases}},
    	author={Marcuzzi, Matteo and Levi, Emanuele and Diehl, Sebastian and Garrahan, Juan P and Lesanovsky, Igor},
    	journal={Phys. Rev. Lett.},
    	volume={113},
    	number={21},
    	pages={210401},
    	year={2014},
    	publisher={APS}
    }

@article{diehl2008quantum,
    	title={Quantum states and phases in driven open quantum systems with cold atoms},
    	author={Diehl, Sebastian and Micheli, Andrea and Kantian, Adrian and Kraus, B and B{\"u}chler, Hans Peter and Zoller, Peter},
    	journal={Nat. Phys.},
    	volume={4},
    	number={11},
    	pages={878--883},
    	year={2008},
    	publisher={Nature Publishing Group UK London}
    }

@article{kraus2008preparation,
    	title={Preparation of entangled states by quantum Markov processes},
    	author={Kraus, Barbara and B{\"u}chler, Hans P and Diehl, Sebastian and Kantian, Adrian and Micheli, Andrea and Zoller, Peter},
    	journal={Phys. Rev. A},
    	volume={78},
    	number={4},
    	pages={042307},
    	year={2008},
    	publisher={APS}
    }

@article{verstraete2009quantum,
    	title={Quantum computation and quantum-state engineering driven by dissipation},
    	author={Verstraete, Frank and Wolf, Michael M and Ignacio Cirac, J},
    	journal={Nat. Phys.},
    	volume={5},
    	number={9},
    	pages={633--636},
    	year={2009},
    	publisher={Nature Publishing Group UK London}
    }

@article{wang2019simulating,
    	title={Simulating Anisotropic quantum {R}abi model via frequency modulation},
    	author={Wang, Gangcheng and Xiao, Ruoqi and Shen, HZ and Sun, Chunfang and Xue, Kang},
    	journal={Sci. Rep.},
    	volume={9},
    	number={1},
    	pages={4569},
    	year={2019},
    	publisher={Nature Publishing Group UK London}
    }

@article{wang2018quantum,
    	title={Quantum criticality and state engineering in the simulated anisotropic quantum {R}abi model},
    	author={Wang, Yimin and You, Wen-Long and Liu, Maoxin and Dong, Yu-Li and Luo, Hong-Gang and Romero, G and You, JQ},
    	journal={New J. Phys.},
    	volume={20},
    	number={5},
    	pages={053061},
    	year={2018},
    	publisher={IOP Publishing}
    }

@article{bamba2016superradiant,
    	title={Superradiant phase transition in a superconducting circuit in thermal equilibrium},
    	author={Bamba, Motoaki and Inomata, Kunihiro and Nakamura, Yasunobu},
    	journal={Phys. Rev. Lett.},
    	volume={117},
    	number={17},
    	pages={173601},
    	year={2016},
    	publisher={APS}
    }

@article{grimsmo2013cavity,
      title={{Cavity-QED} simulation of qubit-oscillator dynamics in the ultrastrong-coupling regime},
      author={Grimsmo, Arne L and Parkins, Scott},
      journal={Phys. Rev. A},
      volume={87},
      number={3},
      pages={033814},
      year={2013},
      publisher={APS}
    }

@article{garcia2015light,
      title={Light-matter decoupling and {$\mathbf{A}^2$} term detection in superconducting circuits},
      author={Garcia-Ripoll, Juan Jose and Peropadre, Borja and De Liberato, Simone},
      journal={Sci. Rep.},
      volume={5},
      number={1},
      pages={16055},
      year={2015},
      publisher={Nature Publishing Group UK London}
    }

@article{puebla2017protected,
      title={Protected ultrastrong coupling regime of the two-photon quantum {Rabi} model with trapped ions},
      author={Puebla, Ricardo and Hwang, Myung-Joong and Casanova, Jorge and Plenio, Martin B},
      journal={Phys. Rev. A},
      volume={95},
      number={6},
      pages={063844},
      year={2017},
      publisher={APS}
    }

@article{ciuti2005quantum,
      title={Quantum vacuum properties of the intersubband cavity polariton field},
      author={Ciuti, Cristiano and Bastard, G{\'e}rald and Carusotto, Iacopo},
      journal={Phys. Rev. B},
      volume={72},
      number={11},
      pages={115303},
      year={2005},
      publisher={APS}
    }

@article{todorov2012intersubband,
      title={Intersubband polaritons in the electrical dipole gauge},
      author={Todorov, Yanko and Sirtori, Carlo},
      journal={Phys. Rev. B},
      volume={85},
      number={4},
      pages={045304},
      year={2012},
      publisher={APS}
    }

@article{gelhausen2018dissipative,
      title={Dissipative {Dicke} model with collective atomic decay: Bistability, noise-driven activation, and the nonthermal first-order superradiance transition},
      author={Gelhausen, Jan and Buchhold, Michael},
      journal={Phys. Rev. A},
      volume={97},
      number={2},
      pages={023807},
      year={2018},
      publisher={APS}
    }

@article{kirton2017suppressing,
      title={Suppressing and restoring the {Dicke} superradiance transition by dephasing and decay},
      author={Kirton, Peter and Keeling, Jonathan},
      journal={Phys. Rev. Lett.},
      volume={118},
      number={12},
      pages={123602},
      year={2017},
      publisher={APS}
    }

@article{beaudoin2011dissipation,
      title={Dissipation and ultrastrong coupling in circuit {QED}},
      author={Beaudoin, F{\'e}lix and Gambetta, Jay M and Blais, A},
      journal={Phys. Rev. A},
      volume={84},
      number={4},
      pages={043832},
      year={2011},
      publisher={APS}
    }

@article{ridolfo2012photon,
      title={Photon blockade in the ultrastrong coupling regime},
      author={Ridolfo, Alessandro and Leib, Martin and Savasta, Salvatore and Hartmann, Michael J},
      journal={Phys. Rev. Lett.},
      volume={109},
      number={19},
      pages={193602},
      year={2012},
      publisher={APS}
    }
\end{document}